\documentclass[letterpaper,twocolumn,twoside]{IEEEtran}
\usepackage{amsmath, amssymb, bm, cite, epsfig, psfrag}
\usepackage{algorithm,algorithmic}
\usepackage{ifthen}
\usepackage{color}
\usepackage{hyperref}
\usepackage[normalem]{ulem} 

\newboolean{anonymous}
\setboolean{anonymous}{false}

\newif\ifarxiv
\arxivtrue 

\newcommand{\textb}[1]{\textcolor{black}{#1}}
\newcommand{\textr}[1]{\textcolor{red}{#1}}

\newcommand{\blue}{\color{black}}

\newcommand{\black}{\color{black}}

\newcommand*\dif{\mathop{}\!\mathrm{d}} 

\def\beq{\begin{equation}}
\def\eeq{\end{equation}}
\def\beqa{\begin{eqnarray}}
\def\eeqa{\end{eqnarray}}
\def\beqan{\begin{eqnarray*}}
\def\eeqan{\end{eqnarray*}}

\def\argmin{\mathop{\mathrm{arg\,min}}}
\def\argmax{\mathop{\mathrm{arg\,max}}}

\def\x{\times}

\def\MID{|}
\def\MIDD{;}

\newtheorem{lemma}{Lemma}

\newtheorem{algorithmEnv}[algorithm]{Algorithm}

\renewcommand{\hat}{\widehat}

\def\phat{\widehat{p}}

\def\rhat{\widehat{r}}
\def\shat{\widehat{s}}

\def\xhat{\widehat{x}}
\def\zhat{\widehat{z}}
\def\la{\leftarrow}
\def\ra{\rightarrow}
\def\LLR{\mbox{\small \sffamily LLR}}

\def\arr{\rightarrow}
\def\Exp{\mathbb{E}}
\def\var{\mathrm{var}}
\def\cov{\mathrm{Cov}}
\def\tm1{t\! - \! 1}
\def\tp1{t\! + \! 1}

\newcommand{\zero}{\mathbf{0}}
\newcommand{\abf}{\mathbf{a}}

\newcommand{\pbf}{\mathbf{p}}
\newcommand{\pbfhat}{\widehat{\mathbf{p}}}

\newcommand{\rbfhat}{\widehat{\mathbf{r}}}
\newcommand{\sbf}{\mathbf{s}}
\newcommand{\sbfhat}{\widehat{\mathbf{s}}}
\newcommand{\ubf}{\mathbf{u}}

\newcommand{\vbf}{\mathbf{v}}

\newcommand{\wbf}{\mathbf{w}}
\newcommand{\wbfhat}{\widehat{\mathbf{w}}}
\newcommand{\xbf}{\mathbf{x}}
\newcommand{\xbfhat}{\widehat{\mathbf{x}}}
\newcommand{\ybf}{\mathbf{y}}
\newcommand{\zbf}{\mathbf{z}}
\newcommand{\zbfhat}{\widehat{\mathbf{z}}}
\newcommand{\Abf}{\mathbf{A}}

\newcommand{\Dbf}{\mathbf{D}}

\newcommand{\Ibf}{\mathbf{I}}

\newcommand{\Qbf}{\mathbf{Q}}

\newcommand{\Vbf}{\mathbf{V}}
\newcommand{\Wbf}{\mathbf{W}}

\newcommand{\Xbf}{\mathbf{X}}
\newcommand{\Xbfhat}{\widehat{\mathbf{X}}}

\def\xibf{{\boldsymbol \xi}}
\def\rhohat{{\widehat{\rho}}}

\newcommand{\bigCond}[2]{\bigl({#1} \!\bigm\vert\! {#2} \bigr)}

\renewcommand{\vec}[1]{\ensuremath{\boldsymbol{#1}}}
\newcommand{\tran}{^{\textsf{T}}}

\def\POPT{\mbox{\small \sffamily P-OPT}}
\def\PEXP{\mbox{\small \sffamily P-EXP} }

\newcommand{\VKG}[1]{\textb{\bf [#1 --Vivek]}}
\newcommand{\PS}[1]{\textr{\bf [#1 --Phil]}}
\renewcommand{\VKG}[1]{}
\renewcommand{\PS}[1]{}

\title{Hybrid Approximate Message Passing
}
\ifthenelse{\boolean{anonymous}}
{
  \author{}
}
{
  \author{Sundeep Rangan, \IEEEmembership{Fellow,~IEEE},
          Alyson K. Fletcher, \IEEEmembership{Member,~IEEE},
          Vivek~K~Goyal, \IEEEmembership{Fellow,~IEEE},\\
          Evan Byrne, \IEEEmembership{Student Member,~IEEE},
          and Philip Schniter, \IEEEmembership{Fellow,~IEEE}%
  \thanks{S. Rangan (email: srangan@nyu.edu) is with
          the Department of Electrical and Computer Engineering, 
          New York University, Brooklyn, NY, 11201\@.
          His work was supported in part by 
          the National Science Foundation under Grant 1116589 
          and the industrial affiliates of NYU WIRELESS.}
  \thanks{A.~K.~Fletcher (email: akfletcher@ucla.edu) is with
          the Department of Statistics and Electrical Engineering, 
          the University of California, Los Angeles, CA, 90095\@.
          Her work was supported in part by 
          the National Science Foundation under Grant 1254204 and 
          the Office of Naval Research under Grant N00014-15-1-2677.}
  \thanks{V.~K. Goyal (email: v.goyal@ieee.org) is with
          the Department of Electrical and Computer Engineering at
          Boston University, Boston, MA, 02215\@.
          His work was supported in part by
          the National Science Foundation under Grant 1422034.}
  \thanks{E.~Byrne and P.~Schniter 
          (email: byrne.133@osu.edu and schniter@ece.osu.edu) are with
          the Department of Electrical and Computer Engineering,
          The Ohio State University, Columbus, OH, 43210\@.
          Their work was supported in part by 
          the National Science Foundation under Grant CCF-1527162.}
  \thanks{Portions of this work were presented at the 
          IEEE International Symposium on Information Theory 
          \cite{RanganFGS:12-ISIT}.}
  \markboth{Hybrid Generalized Approximate Message Passing}
        {Rangan, Fletcher, Goyal, Byrne, and Schniter}
}
}

\begin{document}

\maketitle
\begin{abstract}
Gaussian and quadratic approximations of message passing algorithms on graphs
have attracted considerable recent attention due to their computational simplicity,
analytic tractability, and wide applicability in optimization and
statistical inference problems.
This paper presents a systematic framework for incorporating such approximate
message passing (AMP) methods in general graphical models.
The key concept is a partition of dependencies of a general graphical model
into strong and weak edges, with the weak edges representing
small, linearizable couplings of variables.
AMP approximations based on the Central Limit Theorem can be readily applied to \textb{aggregates of many} weak edges
and integrated with standard message passing updates on the strong edges.
The resulting algorithm, which we call hybrid generalized approximate message passing
(HyGAMP), can yield significantly simpler implementations of sum-product and max-sum
loopy belief propagation.
By varying the partition of strong and weak edges, a performance--complexity trade-off can
be achieved.  
Group sparsity and multinomial logistic regression problems are studied as examples of the proposed methodology.
\end{abstract}



\section{Introduction} \label{sec:intro}

For high-dimensional optimization and inference problems,
message-passing algorithms constructed from graphical models 
have become widely-used in many fields \cite{Frey:98,WainwrightJ:08,Koller:09}.
The fundamental principle of graphical models is to decompose high-dimensional problems
into sets of smaller low-dimensional problems.
The decomposition is represented using a bipartite graph, where the problem variables and
factors are represented by the graph vertices and the dependencies between
them represented by edges.
Message passing methods such as loopy belief propagation (BP)
use this graphical structure
to perform optimization or approximate inference in an iterative manner.
In each iteration, optimization or inference is performed ``locally" on the sub-problems
associated with each factor, and ``messages"
are passed between the variables and factors to
account for the coupling between these sub-problems.

Recently, so-called ``approximate message passing'' (AMP) \cite{DonohoMM:09,DonohoMM:10-ITW1,BayatiM:11} and generalized AMP (GAMP) \cite{Rangan:11-ISIT} methods have been developed for the case where the measurement factors depend weakly on a large number of random variables.
By linearizing these weak dependencies, one can simplify standard loopy-BP algorithms and rigorously analyze their behavior in the high-dimensional limit \cite{BayatiM:11}.
AMP algorithms of this form have been proposed for maximum a posteriori (MAP) and minimum mean-squared error (MMSE) inference in linear models \cite{DonohoMM:09,DonohoMM:10-ITW1}, generalized linear models \cite{Rangan:11-ISIT}, and generalized bilinear models \cite{parker2013bilinear,parker2013bilinear2,parker2016bilinear}.
These AMP algorithms, however, assume that the underlying random variables are independent.
Similarly, they assume that measurements are conditionally independent given these random variables.
Thus, one may wonder how to extend these AMP methods to prior (and/or likelihood) models that include dependencies among variables (and/or measurements).
\textb{By exploiting such dependencies, one can greatly improve the performance of optimization or inference.  (We will show an example of this phenomenon in Section~\ref{sec:grpSparse}.)}


As one solution,
we present \emph{Hybrid GAMP} (HyGAMP) algorithms
for what we call \emph{graphical model problems with linear mixing}.
The basic idea is to partition the edges of the graphical model into 
\emph{weak} and \emph{strong} subsets and represent the dependencies among
the weak edges using a linear transform.
Assuming that the individual components of this linear transform are \textb{individually weak}, 
the messages propagating
on the weak edges can be
simplified using AMP-style approximations and combined 
with standard loopy-BP messages on the strong edges.
The proposed approach is thus a hybrid of AMP and standard loopy-BP techniques.

We detail the HyGAMP methodology using two common variants of loopy BP:
the \emph{sum-product} algorithm for inference (i.e., computation of the posterior mean)
and the \emph{max-sum} algorithm for optimization (i.e., computation of the posterior mode).
For the sum-product loopy BP algorithm, we argue that the weak-edge
messages can be approximated
by Gaussian densities whose mean and variance computations are simplified by the Central
Limit Theorem (CLT).  For max-sum loopy BP, we argue that the weak-edge messages
can use quadratic approximations whose
parameters are easily computed using least-squares techniques.

\blue
The proposed approach can be considered as a generalization of the \emph{turbo AMP} method proposed in \cite{Schniter:10-CISS} for clustered-sparse signal recovery.
The idea behind turbo AMP is to 
i) partition the overall factor graph into sub-graphs with weak edges and sub-graphs with strong edges, 
ii) perform AMP-style message passing within the weak sub-graphs and standard sum-product BP within the strong sub-graphs, and
iii) periodically interchange messages between neighboring sub-graphs.
Although the turbo-AMP idea has been applied to channel estimation and equalization, wavelet image denoising, video compressive sensing, hyperspectral unmixing, and other problems in, e.g., \cite{Schniter:11,SomS:12,ZinielS:13,ZinielS:13b,NassarSE:14,Vila:TCI:15,tramel2016boltzmann}, 
a concrete turbo-AMP algorithm that applies to generic factor graphs has never been stated.
HyGAMP fills this gap.
Furthermore, turbo AMP methods have been proposed exclusively with sum-product message passing.
HyGAMP extends the turbo-AMP idea to max-sum message passing.
Going further still,
the proposed HyGAMP method generalizes turbo-AMP by allowing factor graphs with vector-valued variable nodes (in the strong and/or weak sub-graphs).
As such,
HyGAMP facilitates the application of AMP techniques to problems such as \emph{group-sparse estimation} and \emph{multinomial logistic regression}, which are outside the reach of AMP and turbo AMP.
\black


The use of AMP-style approximations on portions of a factor graph
has also been applied with joint parameter estimation and decoding for CDMA
multiuser detection in \cite{CaireTulBig:01};
in a wireless interference coordination problem in \cite{RanganM:11},
and in the context of compressed sensing \cite[Section 7]{Montanari:12-bookChap}.
The HyGAMP framework presented here unifies and extends all of these examples
and thus provides a systematic procedure for incorporating Gaussian approximations
of message passing in a modular manner in general graphical models.

\blue
A shorter version of this paper was published in \cite{RanganFGS:12-ISIT}.
This longer version includes 
derivations of the proposed algorithms,
additional experiments, and
many additional explanations, clarifications, and examples throughout.
Note that, since the publication of \cite{RanganFGS:12-ISIT}, the HyGAMP methodology
has been used to solve a variety of problems, including
multiuser detection in massive MIMO \cite{wang2015mimo,wang2016signal},
inference for neuronal connectivity \cite{fletcher2014scalable},
fitting neural mass spatio-temporal models \cite{fletcher2015neural},
user activity detection in cloud-radio random access \cite{utkovski2017random},
and
decoding from pooled data \cite{alaoui2017pooled}.
\black

\section{Graphical Model Problems with Linear Mixing} \label{sec:graphModel}
Let $\xbf$ and $\zbf$ be real-valued block column vectors
\beq \label{eq:xzblock}
    \xbf = [\xbf_1\tran, \ldots, \xbf_n\tran]\tran, \qquad
    \zbf = [\zbf_1\tran, \ldots, \zbf_m\tran]\tran,
\eeq
where $\tran$ denotes transposition,
and consider a function of these vectors of the form
\beq \label{eq:Fdef}
    F(\xbf,\zbf) := \sum_{i=1}^m f_i(\xbf_{\alpha(i)}, \zbf_i),
\eeq
where, for each $i$, $f_i(\cdot)$ is a real-valued function;
$\alpha(i)$ is a subset of the indices $\{1,\ldots,n\}$; and
$\xbf_{\alpha(i)}$ is the concatenation of the vectors
$\{ \xbf_j,\, j \in \alpha(i) \}$.
We will be interested in computations on this function
subject to linear constraints of the form
\beq \label{eq:zi}
    \zbf_i = \sum_{j=1}^n \Abf_{ij}\xbf_j = \Abf_i \xbf,
\eeq
where each $\Abf_{ij}$ is a real-valued matrix and $\Abf_i$
is the matrix with block columns $\{\Abf_{ij}\}_{j=1}^n$.
We will also let $\Abf$ be the matrix with block rows $\{\Abf_i\}_{i=1}^m$,
so that we can write the linear constraints simply as $\zbf = \Abf\xbf$.

\begin{figure}
\begin{center}
  \vspace{0.1in}
  \newcommand{\sz}{0.8}
  \newcommand{\szz}{0.6}
  \psfrag{d}[][l][\sz]{$\vdots$}
  \psfrag{x1}[bl][Bl][\sz]{$\xbf_1$}
  \psfrag{x2}[bl][Bl][\sz]{$\xbf_2$}
  \psfrag{x3}[bl][Bl][\sz]{$\xbf_3$}
  \psfrag{xn}[bl][Bl][\sz]{$\xbf_n$}
  \psfrag{f1}[br][Bl][\sz]{$f_1(\xbf_{\alpha(1)},\zbf_1)$}
  \psfrag{f2}[br][Bl][\sz]{$f_2(\xbf_{\alpha(2)},\zbf_2)$}
  \psfrag{fm}[br][Bl][\sz]{$f_m(\xbf_{\alpha(m)},\zbf_m)$}
  \psfrag{A}[][Bl][1.0]{\sf \begin{tabular}{c}
                              $\Abf$\\[-0.5mm]
                              \scriptsize Mixing\\[-1mm]
                              \scriptsize matrix\end{tabular}}
  \psfrag{strong}[][Bl][\szz]{\sf strong edges}
  \psfrag{weak}[][Bl][\szz]{\sf weak edges}
  \includegraphics[width=2.4in]{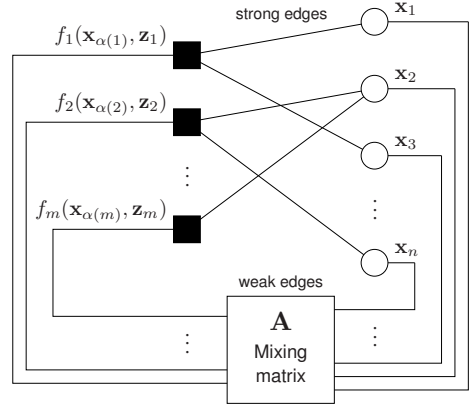}
\end{center}
\caption{Factor graph representation of the linear mixing
estimation and optimization problems.  The variable nodes (circles)
are connected to the factor nodes (squares) either directly
(strong edges) or via the output of the linear mixing matrix $\Abf$ (weak edges).
\textb{The basic GAMP algorithm \cite{Rangan:11-ISIT} handles the special case where there exists no strong edges and where the variables $\xbf_j$ are scalar valued.}}
\label{fig:factorGraph}
\end{figure}

The function $F(\xbf,\zbf)$ is naturally described via a graphical model
as shown in Fig.\ \ref{fig:factorGraph}.
Specifically, we associate with $F(\xbf,\zbf)$
a bipartite \emph{factor graph} $G = (V,E)$ whose
vertices $V$ consist of $n$ \emph{variable nodes}
corresponding to the (vector-valued) variables $\xbf_j$ and $m$ \emph{factor nodes}
corresponding to the factors $f_i(\cdot)$ in \eqref{eq:Fdef}.  There is an edge
$(i,j) \in E$ in the graph if and only if the variable $\xbf_j$ has some influence
on the factor $f_i(\xbf_{\alpha(i)},\zbf_i)$.
This influence can occur in one of two mutually exclusive ways:
\begin{itemize}
\item The index $j$ is in $\alpha(i)$, so that the variable $\xbf_j$
directly appears in the sub-vector $\xbf_{\alpha(i)}$ in the factor
$f_i(\xbf_{\alpha(i)},\zbf_i)$.  In this case, $(i,j)$
will be called a \emph{strong edge}, since $\xbf_j$ can have an arbitrary
and potentially-large influence on the factor.
\item The matrix $\Abf_{ij}$ is nonzero, so that $\xbf_j$ affects
$f_i(\xbf_{\alpha(i)},\zbf_i)$ through its linear
influence on $\zbf_i$ in \eqref{eq:zi}.  In this case, $(i,j)$
will be called a \emph{weak edge}, since the approximations we will
make in the algorithms below assume that 
\textb{$\Abf_{ij}$ are ``small.''}
The set of weak edges into the factor node $i$ will be denoted
$\beta(i)$.
\end{itemize}
\blue
When we say that $\Abf_{ij}$ are ``small,'' we mean
do not mean small in an absolute sense, but rather 
that $\Abf_{ij}$ are such that no individual $\xbf_j$ can have a significant effect 
on the sum $\sum_{j=1}^n \Abf_{ij}\xbf_j$, and likewise that no individual $\zbf_i$ 
can have a significant effect on the sum $\sum_{i=1}^m \zbf_i\tran \Abf_{ij}$.
One example is when $\vec{A}$ is drawn with i.i.d.\ sub-Gaussian entries for 
sufficiently large $m$ and $n$.
Matrices of this type are assumed in derivation and analysis of the AMP methods
\cite{DonohoMM:09,DonohoMM:10-ITW1,BayatiM:11,Rangan:11-ISIT}.
\black

Together, $\alpha(i)$ and $\beta(i)$ comprise the set of all indices $j$
for which a variable node $\xbf_j$ is connected to the factor node
$f_i(\cdot)$ in the graph $G$.  The union $\partial(i) =  \alpha(i) \cup \beta(i)$
is thus the neighbor set of $f_i(\cdot)$.
Similarly, for any variable node
$\xbf_j$, we let (with some abuse of notation)
$\alpha(j)$ be the set of all indices $i$ 
for which a factor node $f_i(\cdot)$ is connected to $\xbf_j$ via a
strong edge, and let $\beta(j)$ be the set of all indices $i$ for which there
exists a weak edge. The union $\partial(j) = \alpha(j) \cup \beta(j)$ 
is thus the neighbor set of $\xbf_j$.

\VKG{The notations $\alpha()$, $\beta()$, and $\partial()$ seem to be
overloaded in an ``unresolvable'' way.  Variables nodes are numbered
$\{1,\,2,\,\ldots,\,n\}$ and factor nodes are numbered $\{1,\,2,\,\ldots,\,m\}$,
so what does $\alpha(1)$ mean?  Does it refer to a factor node (``$i=1$'') or
a variable node (``$j=1$'')?}
\PS{Agreed that it is unresolvable.  Added ``with some abuse of notation'' above.  Would be a lot of work to fix!}

Given these definitions, we are interested in two problems:
\begin{itemize}
\item \textbf{Optimization problem $\POPT$}:  Given a function $F(\xbf,\zbf)$
of the form \eqref{eq:Fdef} and a matrix $\Abf$, compute the maximum:
\beq \label{eq:Fopt}
    \xbfhat = \argmax_{\xbf \,:\, \zbf = \Abf\xbf} F(\xbf,\zbf), \qquad \zbfhat = \Abf\xbfhat.
\eeq
Also, for each $j$, compute the \emph{marginal value} function
\beq \label{eq:margUtil}
    \Delta_j(\xbf_j) := \max_{\xbf_{\backslash j} \,:\, \zbf=\Abf\xbf} F(\xbf,\zbf),
\eeq
where $\xbf_{\backslash j}$ is composed of $\{\xbf_r\}_{r\neq j}$.

\item \textbf{Expectation problem $\PEXP$}:  Given a function $F(\xbf,\zbf)$
of the form \eqref{eq:Fdef}, a matrix $\Abf$, and \emph{scale factor} $u > 0$,
define the joint density
\beq \label{eq:pxz}
    p(\xbf) := Z^{-1}(u) \exp\left[ u F(\xbf, \zbf) \right],
    \qquad \zbf = \Abf\xbf
\eeq
where $Z(u)$ is a normalization constant called the partition function
(which is a function of $u$).
Then, for this density, compute the expectations
\beq \label{eq:xzExp}
     \xbfhat  = \Exp[\xbf], \qquad \zbfhat=\Exp[\zbf].
\eeq
Also, for each $j$, compute the log marginal
\beq \label{eq:margDist}
    \Delta_j(\xbf_j) := \frac{1}{u} \log \int
    \exp\left[uF(\xbf,\zbf)\right] \, \dif\xbf_{\backslash j} .
\eeq
We include the scale factor $u$ so that the definition of $F(\xbf,\zbf)$ 
allows an arbitrary scaling, as in \eqref{eq:Fopt}.
\end{itemize}

We now show that $\POPT$ and $\PEXP$ commonly arise
in statistical inference.  Suppose that we are
given a probability density $p(\xbf)$ of the form \eqref{eq:pxz}
for some function $F(\xbf,\zbf)$.  The function $F(\xbf,\zbf)$
may depend implicitly on some observed vector $\ybf$, so that $p(\xbf)$ represents
the posterior density of $\xbf$ given $\ybf$.  In this context,
the solution $(\xbfhat,\zbfhat)$ to the problem $\POPT$ is precisely
the \emph{maximum a posteriori} (MAP) estimate of $\xbf$ and $\zbf$
given the observations $\ybf$.  Similarly, the solution
 $(\xbfhat,\zbfhat)$ to the problem $\PEXP$ is precisely
the \emph{minimum mean squared error} (MMSE) estimate.
For $\PEXP$,
the function $\Delta_j(\xbf_j)$ is the log marginal density of $\xbf_j$.

\blue
The two problems are related:
A standard large deviations argument \cite{DemboZ:98} shows that,
under suitable conditions,
as $u\arr \infty$ the density $p(\xbf)$ in \eqref{eq:pxz}
concentrates around the maxima $(\xbfhat,\zbfhat)$ in the solution
to the problem $\POPT$.  As a result, the solution $(\xbfhat,\zbfhat)$
to $\PEXP$ converges to the solution to $\POPT$.
\black

\subsection{Further Assumptions and Notation}

In the analysis below, we will assume that,
for each factor node $f_i(\cdot)$, we have that
\beq \label{eq:alphaBetaDist}
    \alpha(i) \cap \beta(i) = \emptyset,
\eeq
i.e., the strong and weak neighbor sets are disjoint.
This assumption introduces no loss of generality:
If an edge $(i,j)$ is both weak and strong,
we can modify the function $f_i(\xbf_{\alpha(i)},\zbf_i)$ to
``move" the influence of $\xbf_j$ from the term $\zbf_i$ into the
direct term $\xbf_{\alpha(i)}$.  For example, suppose that, for some $i$,
\[
    \zbf_i = \Abf_{i1}\xbf_1 + \Abf_{i3}\xbf_3 + \Abf_{i4}\xbf_4 
    \text{~~and~~} \alpha(i) = \{1,2\}.  
\]
In this case, the edge $(i,1)$ is both strong
and weak.  That is, the function $f_i(\xbf_{\alpha(i)},\zbf_i)$ depends
on $\xbf_1$ through both $\xbf_{\alpha(i)}$ and through
$\zbf_i$.  To satisfy assumption \eqref{eq:alphaBetaDist}, we define 
\begin{align*}
    \zbf_i^{\rm new} 
    &= \Abf_{i3}\xbf_3 + \Abf_{i4}\xbf_4 \\
    f_i^{\rm new}(\xbf_{\alpha(i)}, \zbf_i^{\rm new}) 
    &= f_i((\xbf_1,\xbf_2), \Abf_{i1}\xbf_1 + \zbf_i^{\rm new}),
\end{align*}
under which
$f_i(\xbf_{\alpha(i)},\zbf_i) = f_i^{\rm new}(\xbf_{\alpha(i)},\zbf^{\rm new}_i)$.
Thus we can replace $f_i(\cdot)$ and $\zbf_i$ with $f_i^{\rm new}(\cdot)$ and
$\zbf_i^{\rm new}$
that obey \eqref{eq:alphaBetaDist}.

Even when the dependence of a factor $f_i(\xbf_{\alpha(i)},\zbf_i)$
on a variable $\xbf_j$ is only through the linear term $\zbf_i$,
we may still wish to ``move'' the dependence to a strong edge.
The reason is that the HyGAMP algorithm is designed around the assumption
that the linear dependence is weak, i.e., that the elements in $\Abf_{ij}$ are small.
If these elements are not small, then modeling the dependence with a strong
edge improves the accuracy of HyGAMP at the expense of greater computation.

One final notation:  since $\Abf_{ij} \neq \zero$ only when
$j \in \beta(i)$, we may sometimes write the summation \eqref{eq:zi}
as
\beq \label{eq:zibeta}
    \zbf_i = \sum_{j \in \beta(i)}\Abf_{ij}\xbf_j = \Abf_{i,\beta(i)}\xbf_{\beta(i)},
\eeq
where $\xbf_{\beta(i)}$ is the sub-vector of $\xbf$ with components $j \in \beta(i)$
and $\Abf_{i,\beta(i)}$ is the corresponding sub-matrix of 
$\Abf_i$.

\section{Motivating Examples} \label{sec:examples}

We begin with a basic development to show that problems with a
fully separable prior and likelihood fit within our model.
Then we show an extension to more complicated problems.
More detailed examples are deferred to Sections~\ref{sec:grpSparse} and~\ref{sec:mlr}.

\subsubsection*{Linear Mixing and General Output Channel---Independent Sub-Vectors}
As a simple example of a graphical model with linear mixing, consider the
following estimation problem:  An unknown vector $\xbf$ has
independent sub-vectors $\xbf_j$, each with a joint
probability density $p(\xbf_j)$. The vector $\xbf$ is passed through
a linear transform to yield an output $\zbf = \Abf\xbf$.
Each sub-vector $\zbf_i$ then randomly generates
an output $\ybf_i$ with conditional density $p(\ybf_i \MID \zbf_i)$.
The goal is to estimate $\xbf$ given $\Abf$, the observations $\ybf$, and 
knowledge of the densities.

\blue
Common applications of this formulation include the following.
In \emph{compressive sensing} \cite{Montanari:12-bookChap},
$\xbf$ is a sparse vector and $\Abf$ is a sensing matrix.
The measurements $\ybf$ are usually modeled as $\zbf$ plus Gaussian noise,
in which case $p(\ybf_i|\zbf_i)$ is Gaussian.
In \emph{binary linear classification} \cite{Bishop:06}, 
the rows of $\Abf$ are training feature vectors, 
the elements of $\ybf$ are binary training labels, 
and $\xbf$ is a weight vector learned to predict a label from its feature vector.
Here, $p(\ybf_i|\zbf_i)$ is an ``activation function'' that accounts for error in the
linear-prediction model, often based on the logistic sigmoid.
When $n>m$, a sparse weight vector $\xbf$ is sought to avoid over-fitting 
\cite{KrishnapuramCFH:05}.
In \emph{digital communications} settings,
$\xbf$ might be a vector of finite-alphabet symbols and $\Abf$ a matrix representing
the cumulative effect of the modulation, propagation channel, and demodulation 
\cite{CaireTulBig:01}.
Alternatively, $\xbf$ might represent the channel impulse response, in which case $\Abf$ 
is constructed from a training symbol sequence \cite{Schniter:11}.
In either case, $p(\ybf_i|\zbf_i)$ is usually chosen as Gaussian, although a heavy-tailed
distribution can be chosen to model impulsive noise \cite{NassarSE:14}.
\black

\begin{figure}
\begin{center}
  \vspace{0.2in}
  \newcommand{\sz}{0.8}
  \newcommand{\szz}{0.7}
  \newcommand{\szzz}{0.5}
  \psfrag{d}[][l][\sz]{$\vdots$}
  \psfrag{x1}[bl][Bl][\sz]{$\xbf_1$}
  \psfrag{x2}[bl][Bl][\sz]{$\xbf_2$}
  \psfrag{x3}[bl][Bl][\sz]{$\xbf_3$}
  \psfrag{xn}[bl][Bl][\sz]{$\xbf_n$}
  \psfrag{px1}[l][Bl][\sz]{$p(\xbf_1)$}
  \psfrag{px2}[l][Bl][\sz]{$p(\xbf_2)$}
  \psfrag{px3}[l][Bl][\sz]{$p(\xbf_3)$}
  \psfrag{pxn}[l][Bl][\sz]{$p(\xbf_n)$}
  \psfrag{pyz1}[bl][Bl][\szz]{$p(\ybf_1|\zbf_1)$}
  \psfrag{pyz2}[bl][Bl][\szz]{$p(\ybf_2|\zbf_2)$}
  \psfrag{pyzm}[bl][Bl][\szz]{$p(\ybf_m|\zbf_m)$}
  \psfrag{y1}[r][Bl][\sz]{$\ybf_1$}
  \psfrag{y2}[r][Bl][\sz]{$\ybf_2$}
  \psfrag{ym}[r][Bl][\sz]{$\ybf_m$}
  \psfrag{A}[][Bl][1.0]{$\Abf$}
  \psfrag{obsv}[b][Bl][\szzz]{\sf \begin{tabular}{c} output\\ measurements \end{tabular}}
  \psfrag{channel}[b][Bl][\szzz]{\sf \begin{tabular}{c} measurement\\ channels \end{tabular}}
  \psfrag{mixing}[b][Bl][\szzz]{\sf \begin{tabular}{c} mixing\\ matrix \end{tabular}}
  \psfrag{input}[b][Bl][\szzz]{\sf \begin{tabular}{c} input\\ variables \end{tabular}}
  \psfrag{prior}[b][Bl][\szzz]{\sf \begin{tabular}{c} componentwise\\ prior \end{tabular}}
  \includegraphics[width=2.4in]{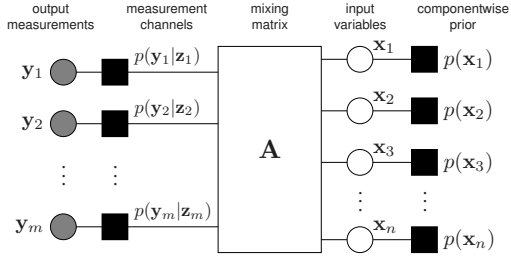}
\end{center}
\caption{An example of a simple graphical model for an estimation problem where
$\xbf$ has independent components with priors $p(\xbf_j)$, $\zbf = \Abf\xbf$,
and the observation vector $\ybf$ is the output of a componentwise measurement
channel with transition function $p(\ybf_i \MID \zbf_i)$. }
\label{fig:graphModSimp}
\end{figure}

Under the assumption that the components $\xbf_j$ are independent and the
components $\ybf_i$ are conditionally independent given $\zbf$, the posterior density on
$\xbf$ factors as
\[
    p(\xbf \MID \ybf) = \frac{1}{Z(\ybf)}\prod_{i=1}^m p(\ybf_i \MID \zbf_i) \prod_{j=1}^n p(\xbf_j), \qquad
    \zbf= \Abf\xbf,
\]
where $Z(\ybf)$ is a normalization constant.
For a fixed observation $\ybf$, we can write this posterior as
\[
    p(\xbf \MID \ybf) \propto \exp\left[ F(\xbf,\zbf)\right], \qquad \zbf = \Abf\xbf,
\]
where $F(\xbf,\zbf)$ is the log posterior, i.e.,
\[
    F(\xbf,\zbf) = \sum_{i=1}^m \log p(\ybf_i \MID \zbf_i) + \sum_{j=1}^n \log p(\xbf_j),
\]
and the dependence on $\ybf$ is implicit.  The log posterior is therefore in the form of
\eqref{eq:Fdef} with scale factor $u=1$ and
$m+n$ factors $\{f_i(\cdot)\}_{i=1}^{m+n}$.
The first $m$ factors can be assigned as 
\begin{align}
    f_i(\zbf_i) = \log p(\ybf_i \MID \zbf_i), \qquad i =1,\ldots,m,
    \label{eq:fi}
\end{align}
which do not directly depend on the terms $\xbf_j$.  Thus $\alpha(i) = \emptyset$ for each $i=1,\dots,m$.
The remaining $n$ factors are then 
\begin{align}
    f_{m+j}(\xbf_j) = \log p(\xbf_j), \qquad j =1,\ldots,n.
    \label{eq:fmj}
\end{align}
For these factors, the strong edge set is the singleton $\alpha(m+j) = \{j\}$
for $j=1,\dots,n$,
and there is no linear term; we can think of $\{\zbf_{m+j}\}_{j=1}^n$
as zero-dimensional.
The corresponding factor graph with the $m+n$ factors
is shown in Fig.~\ref{fig:graphModSimp}.

In the case when all $\xbf_j$ and $\zbf_i$ are scalars, the estimation
problem is precisely the 
one targeted by GAMP \cite{Rangan:11-ISIT},
as mentioned in the introduction.  The special subcase 
of measurements in additive white Gaussian noise (AWGN), i.e.,
\beq \label{eq:yzwGauss}
    y_i = z_i + w_i, \qquad w_i \sim {\cal N}(0,\sigma^2_w),
\eeq
is the one targeted by AMP \cite{DonohoMM:09,DonohoMM:10-ITW1,BayatiM:11}.


\subsubsection*{Linear Mixing and General Output Channel---Dependent Sub-Vectors}
\begin{figure}
\begin{center}
  \newcommand{\sz}{0.8}
  \newcommand{\szz}{0.6}
  \newcommand{\szzz}{0.5}
  \psfrag{.}{}
  \psfrag{d}[][l][\sz]{$\vdots$}
  \psfrag{x1}[bl][Bl][\szz]{$\xbf_1$}
  \psfrag{x2}[bl][Bl][\szz]{$\xbf_2$}
  \psfrag{x3}[bl][Bl][\szz]{$\xbf_3$}
  \psfrag{xn}[bl][Bl][\szz]{$\xbf_n$}
  \psfrag{u1}[l][Bl][\sz]{$\ubf_1$}
  \psfrag{u2}[l][Bl][\sz]{$\ubf_2$}
  \psfrag{u3}[l][Bl][\sz]{$\ubf_k$}
  \psfrag{px1}[l][Bl][\szz]{$p(\xbf_1|\ubf)$}
  \psfrag{px2}[l][Bl][\szz]{$p(\xbf_2|\ubf)$}
  \psfrag{px3}[l][Bl][\szz]{$p(\xbf_3|\ubf)$}
  \psfrag{pxn}[l][Bl][\szz]{$p(\xbf_n|\ubf)$}
  \psfrag{pyz1}[bl][Bl][\szz]{$p(\ybf_1|\zbf_1,\vbf)$}
  \psfrag{pyz2}[bl][Bl][\szz]{$p(\ybf_2|\zbf_2,\vbf)$}
  \psfrag{pyzm}[bl][Bl][\szz]{$p(\ybf_m|\zbf_m,\vbf)$}
  \psfrag{y1}[r][Bl][\sz]{$\ybf_1$}
  \psfrag{y2}[r][Bl][\sz]{$\ybf_2$}
  \psfrag{ym}[r][Bl][\sz]{$\ybf_m$}
  \psfrag{v1}[r][Bl][\sz]{$\vbf_1$}
  \psfrag{v2}[r][Bl][\sz]{$\vbf_2$}
  \psfrag{vm}[r][Bl][\sz]{$\vbf_m$}
  \psfrag{A}[][Bl][1.0]{$\Abf$}
  \psfrag{outpar}[b][Bl][\szzz]{\sf \begin{tabular}{c} output\\ parameters \end{tabular}}
  \psfrag{obsv}[b][Bl][\szzz]{\sf \begin{tabular}{c} output\\ measurements \end{tabular}}
  \psfrag{channel}[b][Bl][\szzz]{\sf \begin{tabular}{c} measurement\\ channels \end{tabular}}
  \psfrag{mixing}[b][Bl][\szzz]{\sf \begin{tabular}{c} mixing\\ matrix \end{tabular}}
  \psfrag{input}[b][Bl][\szzz]{\sf \begin{tabular}{c} input\\ variables \end{tabular}}
  \psfrag{prior}[b][Bl][\szzz]{\sf \begin{tabular}{c} componentwise\\ prior \end{tabular}}
  \psfrag{inpar}[b][Bl][\szzz]{\sf \begin{tabular}{c} input\\ parameters \end{tabular}}
  \includegraphics[width=3.4in]{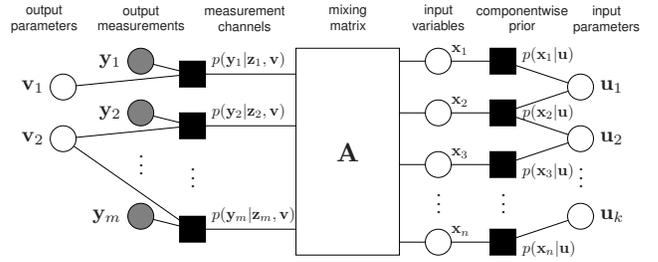}
\end{center}
\caption{A generalization of the model in Fig.\ \ref{fig:graphModSimp}, where the
input variables $\xbf$ are themselves generated by a graphical model with latent variables $\ubf$.
Similarly, the dependence of the observation vector $\ybf$ on the linear mixing output $\zbf$
is through a second graphical model.}
\label{fig:graphModTurbo}
\end{figure}

We now consider the significantly more general graphical model framework 
shown in Fig.~\ref{fig:graphModTurbo}.
In this case, the input sub-vectors $\xbf_j$ may be statistically dependent
on one another, with dependences described by a graphical model.  
Some additional latent variables, in a vector $\ubf$, may also be involved.
For example,
\cite{Schniter:10-CISS} used a discrete Markov chain to model clustered
sparsity, \cite{ZinielS:13b} used discrete-Markov and Gauss-Markov chains to model slow changes in support and amplitude across
multiple measurement vectors, and \cite{SomS:12} used a discrete Markov
tree to model persistence across scale in the wavelet
coefficients of an image.
In Section~\ref{sec:grpSparse}, we will detail the application of HyGAMP to 
group sparsity.

Similarly, the likelihood need not be separable in $\{\ybf_i\}$.
For example, the observations $\ybf_i$ can depend on the outputs $\zbf_i$ through a second graphical model that may include additional latent variables $\vbf_i$.
For example, the distribution of $\ybf_1$ given $\zbf_1$ may depend on unknown parameters $\vbf$ that also affect the distribution of $\ybf_2$ given $\zbf_2$. 
This technique was used in \cite{Schniter:11} to incorporate constraints on LDPC coded bits when performing turbo sparse-channel estimation, equalization, and decoding using GAMP\@.
In Section~\ref{sec:mlr}, we will detail the application of HyGAMP to 
multinomial logistic regression.

\section{Review of Loopy Belief Propagation} \label{sec:stdBP}

Finding exact solutions to high-dimensional $\POPT$ and $\PEXP$ problems 
is generally intractable
because they require
optimization or expectation over
$n$ variables $\xbf_j$.
A widely-used approximation method is loopy BP 
\cite{Pearl:88,WainwrightJ:08}, which reduces the high-dimensional
problem to a sequence of low-dimensional problems
associated with each factor $f_i(\xbf_{\alpha(i)},\zbf_i)$.
We consider two common variants of loopy BP:
the max-sum algorithm (MSA) for the problem $\POPT$ and the
sum-product algorithm (SPA) for the problem $\PEXP$.
This section will briefly review these methods,
as they will be the basis of
the HyGAMP algorithms described in Section~\ref{sec:gampAlgo}.

The MSA iteratively passes
estimates of the marginal utilities
$\Delta_j(\xbf_j)$ in \eqref{eq:margUtil} along the graph edges.
Similarly, the SPA passes estimates of the
log marginals $\Delta_j(\xbf_j)$ in \eqref{eq:margDist}.
For either algorithm, we index the iterations by $t=0,1,2,...$
and denote the ``message" from the factor node $f_i$ to
the variable node $\xbf_j$ in the $t$th iteration by $\Delta_{i \ra j}(t,\xbf_j)$
and the reverse message by $\Delta_{i \la j}(t,\xbf_j)$.

To describe the message updates, we introduce some additional notation.
First, we note that SPA and MSA messages are equivalent up to a constant offset.
That is, adding any constant (w.r.t.\ $\xbf_j$) to either 
$\Delta_{i \ra j}(t,\xbf_j)$ or $\Delta_{i \la j}(t,\xbf_j)$ has no effect on the algorithm.
Thus, we will use ``$\equiv$'' for equality up to a constant offset, i.e.,
\[
    \Delta(\xbf) \equiv g(\xbf) ~\Leftrightarrow~
    \Delta(\xbf) = g(\xbf) + C,
\]
for some constant $C$ that does not depend on $\xbf$.  Similarly, we write
$p(\xbf) \propto q(\xbf)$ when $p(\xbf) = Cq(\xbf)$ for some
constant $C$.  Finally, for the SPA,
we will fix the scale factor $u>0$ in the problem $\PEXP$, and, for any
function $\Delta(\cdot)$, we will write $\Exp[g(\xbf) \MIDD \Delta(\cdot)]$ to denote
the expectation of $g(\xbf)$ with respect to the density $p(\xbf)$ associated with 
$\Delta(\cdot)$:
\begin{align}
    \Exp[g(\xbf) \MIDD \Delta(\cdot)] &= \int g(\xbf)p(\xbf) \, \dif\xbf
    \label{eq:expDel} \\
    p(\xbf) &\propto \exp\left[u\Delta(\xbf)\right]
\end{align}
Given these definitions, the updates for the MSA and SPA variants of
loopy BP are as follows:

\smallskip

\begin{algorithmEnv} \label{algo:LBP}
\textbf{Loopy BP:}
Consider the problems $\POPT$ or $\PEXP$ above for some
function $F(\xbf,\zbf)$ of the form \eqref{eq:Fdef} and
matrix $\Abf$.  For the problem $\PEXP$, fix the scale factor $u > 0$.
The \textbf{MSA} for $\POPT$ and
the \textbf{SPA} for $\PEXP$ iterate the
following steps:
\end{algorithmEnv}
\begin{enumerate}
\item[0)] \emph{Initialization:}  Set $t=0$ and, for each $(i,j) \in E$, set
$\Delta_{i \la j}(t,\xbf_j) = 0$.

\item[1)] \emph{Factor node update:} For each edge $(i,j) \in E$, compute
the function
\beqa
    \lefteqn{ H_{i \ra j}(t,\xbf_{\partial(i)},\zbf_i) } \nonumber \\
    &:=&
    f_i(\xbf_{\alpha(i)}, \zbf_i) + \sum_{r \in \{\partial(i) \setminus j\}}
        \Delta_{i \la r}(t,\xbf_r).
     \label{eq:HijBp}
\eeqa
For the \textbf{MSA}, compute:
\beq \label{eq:DelijFnBp}
    \Delta_{i \ra j}(t,\xbf_j)  \equiv
    \max_{\substack{\xbf_{\partial(i) \backslash j} \\
    \zbf_i=\Abf_i\xbf}} H_{i \ra j}(t,\xbf_{\partial(i)},\zbf_i),
\eeq
where the maximization is over all variables $\xbf_r$ with $r \in \partial(i)\setminus j$
and subject to the constraint $\zbf_i = \Abf_i\xbf$.

For the \textbf{SPA}, compute:
\beq \label{eq:DelijFnBpSp}
    \Delta_{i \ra j}(t,\xbf_j)  \equiv \frac{1}{u} \log \int
    p_{i \ra j}(t,\xbf_{\partial(i)}) \, \dif\xbf_{\partial(i) \backslash j},
\eeq
where the integration is over all variables $\xbf_r$ with $r \in \partial(i)\setminus j$, and
$p_{i \ra j}(t,\xbf_{\partial(i)})$ is the probability density
\beqa \label{eq:pijFnBpSp}
    p_{i \ra j}(t,\xbf_{\partial(i)}) \propto
    \exp\left[uH_{i \ra j}(t,\xbf_{\partial(i)},\zbf_i=\Abf_i\xbf)\right].
\eeqa

\item[2)] \emph{Variable node update:} For each $(i,j) \in E$:
\beq \label{eq:DelijVarBp}
    \Delta_{i \la j}(\tp1,\xbf_j) \equiv \sum_{\ell \in \{\partial(j) \setminus i\}}
    \Delta_{\ell \ra j}(t,\xbf_j).
\eeq
Also, let
\beq \label{eq:DeljVarBp}
    \Delta_{j}(\tp1,\xbf_j) \equiv \sum_{i \in \partial(j)} \Delta_{i \ra j}(t,\xbf_j).
\eeq
For the \textbf{MSA}, compute:
\beq \label{eq:xhatjBp}
    \xbfhat_j(\tp1) := \argmax_{\xbf_j}\Delta_{j}(\tp1,\xbf_j).
\eeq
For the \textbf{SPA}, compute:
\beq \label{eq:xhatjBpSp}
    \xbfhat_j(\tp1) := \Exp\left[ \xbf_j \MIDD \Delta_{j}(\tp1,\cdot)\right].
\eeq

\item[3)] Increment $t$ and return to Step~1 unless a maximum number of iterations is exceeded.
\end{enumerate}

When the graph $G$ is
acyclic, it can be shown that the MSA and SPA algorithms above converge 
to the exact solutions to the 
$\POPT$ and $\PEXP$ problems, respectively.
When the graph $G$ has cycles, however, the above algorithms are---in general---only approximate, but often quite accurate.
\textb{The previous two statements assume that the loopy-BP messages are computed exactly, which is feasible when all variables are either Gaussian or discrete, but otherwise difficult---incurring a complexity that is exponential in general.}
For more details on loopy BP, see
\cite{Pearl:88,WainwrightJ:08,YedidiaFW:03}.


\section{Hybrid GAMP} \label{sec:gampAlgo}
The HyGAMP algorithm modifies loopy BP  
by replacing the weak edges with approximations of their cumulative effects.
By treating a subset of $d$ dependencies as weak (as in AMP) rather than strong (as in loopy BP), the complexity of handling those dependencies shrinks from exponential in $d$ to linear in $d$.
%
%
In particular, HyGAMP assumes 
the elements of $\Abf_{ij}$ are small along any weak edge $(i,j)$.
Under this assumption,
MSA-HyGAMP uses
a quadratic approximation of the messages along the weak edges,
reducing the factor-node update to a standard least-squares problem.
Similarly, SPA-HyGAMP uses
a Gaussian approximation of the weak-edge messages and applies the CLT
at the factor nodes.

A derivation of the HyGAMP algorithm is given in
Appendix~\ref{sec:SumProdDeriv} for the SPA 
\textb{and Appendix~\ref{sec:MaxSumDeriv} for the MSA\@.}
We note that these derivations are ``heuristic''
in the sense that we do not claim any formal matching between loopy BP and the
HyGAMP approximation.

To state the HyGAMP algorithm, we need additional notation.
At iteration $t$, the HyGAMP algorithm produces estimates $\xbfhat_j(t)$ and $\zbfhat_i(t)$
of the vectors $\xbf_j$ and $\zbf_i$.  Several other intermediate vectors,
$\pbfhat_i(t)$, $\sbfhat_i(t)$ and $\rbfhat_j(t)$, are also produced. Associated with
each of these vectors are matrices like $\Qbf^x_j(t)$ and $\Qbf^z_i(t)$ that represent
Hessians for the MSA and covariances for the SPA\@.
When referring to the inverses of these matrices, we use the notation 
$\Qbf^{-x}_j(t)$ to mean $(\Qbf^x_j(t))^{-1}$.
Finally, for any positive definite matrix $\Qbf$ and vector $\abf$, we
define
$\|\abf\|^2_{\Qbf} := \abf\tran\Qbf^{-1}\abf$, which is a weighted two norm.

\smallskip

\begin{algorithmEnv} \label{algo:GAMP}
\textbf{HyGAMP:}
Consider the problem $\POPT$ or $\PEXP$ for some
function $F(\xbf,\zbf)$ of the form \eqref{eq:Fdef} and
matrix $\Abf$.  For the problem $\PEXP$, fix the scale factor $u > 0$.  The
\textbf{MS-HyGAMP} algorithm for $\POPT$ and
the \textbf{SP-HyGAMP} algorithm for $\PEXP$ iterate the
following steps:
\end{algorithmEnv}
\begin{enumerate}
\item[0)] \emph{Initialization:}
Set $t=0$ and $\sbfhat_i(-1)=\zero~\forall i$, and
select some initial values $\Delta_{i \ra j}(-1,\xbf_j)$ for each strong
edge $(i,j)$, and $\rbfhat_j(-1)$ and $\Qbf^r_j(-1)$ for each
index $j$.

\item[1)] \emph{Variable node update, strong edges:}  For each strong edge $(i,j)$,
compute
\beqa
 \lefteqn{ \Delta_{i \la j}(t,\xbf_j)
       \equiv \sum_{\ell \in \{ \alpha(j) \setminus i \}}
    \Delta_{\ell \ra j}(\tm1,\xbf_j)  } \nonumber \\
        & & \qquad \qquad - \ \frac{1}{2}\|\rbfhat_j(\tm1)-\xbf_j\|^2_{\Qbf^r_j(\tm1)}. \label{eq:DelVarStri}
\eeqa

\item[2)] \emph{Variable node update, weak edges:}  For each variable node $j$,
compute
\beq \label{eq:DeljVar}
    \Delta_j(t,\xbf_j) \equiv H^x_j(t,\xbf_j,\rbfhat_j(\tm1),\Qbf^r_j(\tm1))
\eeq
and
\beqa
    \lefteqn{ H^x_j(t,\xbf_j,\rbfhat_j,\Qbf^r_j) } \nonumber \\
       &=& \sum_{i \in \alpha(j)} \Delta_{i \ra j}(\tm1,\xbf_j)
        - \frac{1}{2}\|\rbfhat_j-\xbf_j\|^2_{\Qbf^r_j}. \quad \label{eq:Hxj}
\eeqa
For \textbf{MS-HyGAMP},
\begin{subequations} \label{eq:xupdate}
\beqa
    \xbfhat_j(t) &=& \argmax_{\xbf_j}  \Delta_{j}(t,\xbf_j) 
    \label{eq:xupdate1}\\
    \Qbf^{-x}_j(t) &=& -\frac{\partial^2}{\partial \xbf^2}
        \Delta_{j}(t,\xbf_j)
    \label{eq:xupdate2} .
\eeqa
\end{subequations}
For \textbf{SP-HyGAMP},
\begin{subequations} \label{eq:xupdateSp}
\beqa
    \xbfhat_j(t) &=& \Exp\left( \xbf_j \MIDD \Delta_{j}(t,\cdot) \right) 
    \label{eq:xupdateSp1} \\
    \Qbf^{x}_j(t) &=& u\,\cov\left( \xbf_j \MIDD \Delta_{j}(t,\cdot) \right)
    \label{eq:xupdateSp2} .
\eeqa
\end{subequations}

\item[3)] \emph{Factor node update, linear step:}  For each factor node $i$, compute
\begin{subequations} \label{eq:pupdate}
\beqa
    \zbfhat_{i}(t) &=& \sum_{j \in \beta(i)} \Abf_{ij}\xbfhat_{j}(t)
  \label{eq:pupdate_a} \\
    \pbfhat_{i}(t) &=& \zbfhat_i(t) - \Qbf^p_{i}(t)\sbfhat_i(\tm1)
  \label{eq:pupdate_b} \\
    \Qbf^p_{i}(t) &=& \sum_{j \in \beta(i)}
        \Abf_{ij}\Qbf^x_j(t)\Abf_{ij}\tran \label{eq:Qpupdate} .
\eeqa
\end{subequations}

\item[4)] \emph{Factor node update, strong edges:}
For each strong edge $(i,j)$, compute:
\beqa
    \lefteqn{ H^z_{i \ra j}(t,\xbf_{\alpha(i)},\zbf_i,\pbfhat_i,\Qbf^p_i)
    := f_i(\xbf_{\alpha(i)}, \zbf_i)
    } \nonumber \\
    &+& \sum_{r \in \{\alpha(i) \setminus j\}}  \Delta_{i \la r}(t,\xbf_r)
    - \frac{1}{2} \|\zbf_i - \pbfhat_i \|^2_{\Qbf^p_i}.
    \label{eq:Hquadij}
\eeqa
Then, for \textbf{MS-HyGAMP}, compute:
\beqa
    \lefteqn{ \Delta_{i \ra j}(t,\xbf_j) } \nonumber \\
    &=& \max_{\xbf_{\alpha(i) \backslash j}, \zbf_i}
    H^z_{i \ra j}(t,\xbf_{\alpha(i)},\zbf_i,\pbfhat_i(t),\Qbf^p_i(t)),
\quad
    \label{eq:DelijFnOpt}
\eeqa
where the maximization is jointly over $\zbf_i$ and
all components $\xbf_r$ with $r \in \{\alpha(i) \setminus j\}$.

For \textbf{SP-HyGAMP}, compute:
\beq \label{eq:DelijFnSp}
    \Delta_{i \ra j}(t,\xbf_j) \equiv \frac{1}{u} \log  \int
    p_{i \ra j}(t,\xbf_{\alpha(i)},\zbf_i) \dif\xbf_{\alpha(i) \backslash j} \dif\zbf_i ,
\eeq
where the integral is over $\zbf_i$ and
all components $\xbf_r$ with $r \in \{\alpha(i) \setminus j\}$, and
$p_{i \ra j}(t,\xbf_j)$ is the probability density
\beqa
    \lefteqn{ p_{i \ra j}(t,\xbf_{\alpha(i)}, \zbf_i) \propto } \nonumber\\
    & & \exp\left(uH^z_{i \ra j}(t,\xbf_{\alpha(i)}.
    \zbf_i,\pbfhat_i(t),\Qbf^p_i(t))\right). \label{eq:pijFnSp}
\eeqa

\item[5)] \emph{Factor node update, weak edges:}
For each factor node $i$, compute
\begin{subequations} \label{eq:xzDmax}
\beqa
    \lefteqn{ H^z_{i}(t,\xbf_{\alpha(i)},\zbf_i,\pbfhat_i,\Qbf^p_i)
    := f_i(\xbf_{\alpha(i)}, \zbf_i)
    } \nonumber \\
    &+& \sum_{r \in \alpha(i)}  \Delta_{i \la r}(t,\xbf_r)
    - \frac{1}{2} \|\zbf_i - \pbfhat_i \|^2_{\Qbf^p_i}.
    \label{eq:Hquadi}
\eeqa
Then, for \textbf{MS-HyGAMP}, compute:
\beqa
    \lefteqn{ (\xbfhat^0_{\alpha(i)}(t), \zbfhat^0_i(t))  }
       \nonumber \\
      &:=& \argmax_{\xbf, \zbf_i}
       H^z_i(t,\xbf_{\alpha(i)},\zbf_i,\pbfhat_i(t),\Qbf^p_i(t)),
      \hspace{0.2in} \label{eq:xz0max}   \\
       \lefteqn{ \Dbf^z_i(t) :=
    -\frac{\partial^2}{\partial \zbf_i^2}
        H^z_{i}(t,\xbfhat^0_{\alpha(i)},\zbfhat^0_i,\pbfhat_i(t),\Qbf^p_i(t)),
       \label{eq:Dz}}
\eeqa
\end{subequations}
where the maximization in \eqref{eq:xz0max}
is over the sub-vector $\xbf_{\alpha(i)}$ and
output vector $\zbf_i$.

For \textbf{SP-HyGAMP}, let
\beq \label{eq:zDsp}
    \zbfhat^0_i(t) = \Exp(\zbf_i), \qquad \Dbf^{-z}_i(t) = u\,\cov(\zbf_i),
\eeq
where $\zbf_i$ is the component of the pair $(\xbf_{\alpha(i)},\zbf_i)$
with the joint density
\beqa
    \lefteqn{ p_{i}(t,\xbf_{\alpha(i)}, \zbf_i) \propto } \nonumber\\
    & & \exp\left(uH^z_{i}(t,\xbf_{\alpha(i)},
    \zbf_i,\pbfhat_i(t),\Qbf^p_i(t))\right). \label{eq:piFnSp}
\eeqa
Then, for either MS-HyGAMP or SP-HyGAMP compute
\begin{subequations} \label{eq:supdate}
\beqa
    \sbfhat_i(t) &=& \Qbf^{-p}_i(t)\left[ \zbfhat_i^0(t)-\pbfhat_i(t)\right], \\
    \Qbf^s_i(t) &=& \Qbf^{-p}_i(t)-\Qbf^{-p}_i(t)\Dbf^{-z}_i(t)\Qbf^{-p}_i(t).
\quad
\eeqa
\end{subequations}

\item[6)] \emph{Variable node update, linear step:}  For each variable node $j$
compute
\begin{subequations} \label{eq:rupdate}
\beqa
    \Qbf^{-r}_{j}(t) &=& \sum_{i \in \beta(j)}
        \Abf_{ij}\tran\Qbf^s_i(t)\Abf_{ij} \label{eq:Qrupdate}, \label{eq:rupdate_a} \\
    \rbfhat_{j}(t) &=&  \xbfhat(t) + \Qbf^{r}_{j}(t)
        \sum_{i \in \beta(j)} \Abf_{ij}\tran\sbfhat_i(t). \label{eq:rupdate_b}
\eeqa
\end{subequations}
Increment $t$ and return to Step~1 unless either a maximum number of iterations is exceeded or $\|\xbfhat_j(t)-\xbfhat_j(\tm1)\|$ is sufficiently small.
\end{enumerate}

Although the HyGAMP algorithm above appears much more complicated than
standard loopy BP (Algorithm \ref{algo:LBP}), HyGAMP can require dramatically less computation.
Recall that the main computational difficulty of loopy BP is Step~1, the factor update.
The updates \eqref{eq:DelijFnBp} and \eqref{eq:DelijFnBpSp}
involve an optimization or expectation over $|\partial(i)|$ sub-vectors,
where $\partial(i)$ is the set of all sub-vectors connected to the factor node $i$.
In the HyGAMP algorithm, these computations are replaced by \eqref{eq:DelijFnOpt}
and \eqref{eq:DelijFnSp}, where the optimization and expectation need only be computed
over the strong edge sub-vectors $\alpha(i)$.  If the number of weak edges is large, the computational
savings can be dramatic.  The other steps of the HyGAMP algorithms are all linear, simple
least-square operations, or componentwise nonlinear functions on the individual sub-vectors.

\blue
For ease of illustration, we have only presented one form of the HyGAMP procedure.
Several variants are possible:
\begin{itemize}
\item \emph{Discrete distributions:}  The above description assumed continuous-valued
random variables $x_j$.  The procedures can be easily modified for
discrete-valued variables by appropriately replacing integrals with summations.

\item \emph{Message scheduling:}  The above description also only considered a
completely parallel implementation
 where each iteration performs exactly one update on all edges.
Other so-called message schedules are also possible and may offer more efficient
implementations or better convergence depending on the application
(e.g., \cite{MalioutovJW:06,ElidanMK:06,manoel2015swamp}).
\end{itemize}
\black

\section{Application to Group-Sparse Signal Recovery} \label{sec:grpSparse}

To illustrate the HyGAMP method, we first consider the \emph{group-sparse estimation} problem
\cite{YuanLin:06,ZhaoRY:08}.
Although this problem does not utilize the full generality of HyGAMP,
it provides a simple example of the HyGAMP method and
has a number of existing algorithms that can be compared against.

\subsection{HyGAMP Algorithm}

A general version of the group-sparsity problem that falls within the HyGAMP
framework can be described as follows.  Let $\xbf$ be an $n$-dimensional
vector with scalar components $\{x_j\}_{j=1}^n$.
Vector-valued components could also be considered, but we restrict our attention to
scalar components for simplicity.
The component indices $j$ of the vector $\xbf$ are
divided into $K$ (possibly overlapping) groups, $G_1,\ldots,G_K \subseteq \{1,\ldots,n\}$.
We let $\gamma(j)$ be the set of group indices $k$ such that $j \in G_k$.
That is, $\gamma(j)$ is the set of groups to which the component $x_j$
belongs.

Suppose that each group $G_k$ can be ``active" or ``inactive", and each
component $x_j$ can be non-zero only when at least one group $G_k$ is active
for some $k \in \gamma(j)$.  Qualitatively, a vector $\xbf$ is sparse with respect
to this group structure if it is consistent with only a small number of groups
being active.  That is, most of the components of $\xbf$ are zero with the non-zero
components having support contained in a union of a small number of groups.
The group-sparse estimation problem is to estimate the vector $\xbf$ from some
measurements $\ybf$.  The traditional (non-group) sparse estimation problem
corresponds to the special case when there are $n$ groups of singletons, $G_j = \{j\}$.

There are many ways to
model the group-sparse structure in a Bayesian manner, 
particularly with overlapping groups.
For sake of illustration, we consider
the following simple model.  For each group $G_k$, let
$\xi_k \in \{0,1\}$ be a Boolean variable with $\xi_k=1$ when the group $G_k$
is active and $\xi_k=0$ when it is inactive.
We call $\xi_k$ the ``activity indicators" and model them as i.i.d.\ with
\beq \label{eq:xiProb}
    P(\xi_k=1) = 1-P(\xi_k=0) = \rho
\eeq
for some sparsity rate $\rho \in (0,1)$.  We assume that, given the vector $\xibf$,
the components of $\xbf$ are independent with the conditional densities
\beq\label{eq:xxi}
    x_j|\xibf \sim \left\{ \begin{array}{cl}
        0 & \mbox{if }\xi_k=0 \mbox{ for all } k \in \gamma(j) \\
        V & \mbox{otherwise},
        \end{array} \right.
\eeq
where $V$ is a random variable having the distribution of the component
$x_j$ in the event that it belongs to an active group.
Finally, suppose that measurement vector $\ybf$
is generated by first passing $\xbf$ through a linear transform $\zbf=\Abf\xbf$
and then a separable componentwise measurement channel with likelihoods
$p(y_i|z_i)$.
Many other dependencies
on the activities of $\xbf$ and measurement models $\ybf$ are possible
-- we use this simple model for illustration.

\begin{figure}
\begin{center}
  \newcommand{\sz}{0.8}
  \newcommand{\szz}{0.75}
  \newcommand{\szzz}{0.5}
  \psfrag{.}{}
  \psfrag{d}[][l][\sz]{$\vdots$}
  \psfrag{x1}[bl][Bl][\szz]{$\xbf_1$}
  \psfrag{x2}[bl][Bl][\szz]{$\xbf_2$}
  \psfrag{x3}[bl][Bl][\szz]{$\xbf_3$}
  \psfrag{xn}[bl][Bl][\szz]{$\xbf_n$}
  \psfrag{u1}[bl][Bl][\sz]{$\xi_1$}
  \psfrag{uk}[bl][Bl][\sz]{$\xi_K$}
  \psfrag{px1}[l][Bl][0.65]{$p(\xbf_1|\xibf_{\gamma(1)})$}
  \psfrag{px2}[l][Bl][0.65]{$p(\xbf_2|\xibf_{\gamma(2)})$}
  \psfrag{px3}[l][Bl][0.65]{$p(\xbf_3|\xibf_{\gamma(3)})$}
  \psfrag{pxn}[l][Bl][0.65]{$p(\xbf_n|\xibf_{\gamma(n)})$}
  \psfrag{pu1}[l][Bl][\szz]{$p(\xi_1)$}
  \psfrag{puk}[l][Bl][\szz]{$p(\xi_K)$}
  \psfrag{pyz1}[bl][Bl][\szz]{$p(\ybf_1|\zbf_1)$}
  \psfrag{pyz2}[bl][Bl][\szz]{$p(\ybf_2|\zbf_2)$}
  \psfrag{pyzm}[bl][Bl][\szz]{$p(\ybf_m|\zbf_m)$}
  \psfrag{y1}[r][Bl][\sz]{$\ybf_1$}
  \psfrag{y2}[r][Bl][\sz]{$\ybf_2$}
  \psfrag{ym}[r][Bl][\sz]{$\ybf_m$}
  \psfrag{A}[][Bl][1.0]{$\Abf$}
  \psfrag{obsv}[b][Bl][\szzz]{\sf \begin{tabular}{c} output\\ measurements \end{tabular}}
  \psfrag{channel}[b][Bl][\szzz]{\sf \begin{tabular}{c} measurement\\ channels \end{tabular}}
  \psfrag{mixing}[b][Bl][\szzz]{\sf \begin{tabular}{c} mixing\\ matrix \end{tabular}}
  \psfrag{input}[b][Bl][\szzz]{\sf \begin{tabular}{c} input\\ variables \end{tabular}}
  \psfrag{prior}[b][Bl][\szzz]{\sf \begin{tabular}{c} componentwise\\ prior \end{tabular}}
  \psfrag{inpar}[b][Bl][\szzz]{\sf \begin{tabular}{c} input\\ parameters \end{tabular}}
  \psfrag{inpri}[b][Bl][\szzz]{\sf \begin{tabular}{c} parameter\\ prior \end{tabular}}
  \includegraphics[width=3.3in]{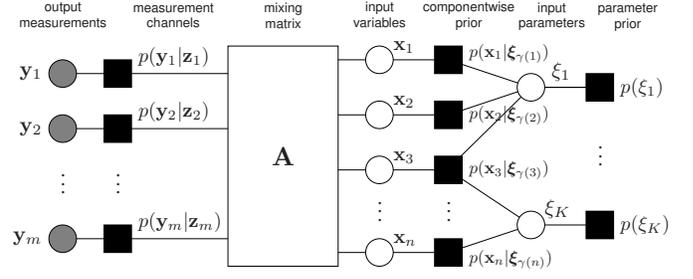}
\end{center}
\caption{Graphical model for the group sparsity problem
with overlapping groups.
The group dependencies between components of the vector $\xbf$ are
modeled via a set of binary latent variables $\xibf$. }
\label{fig:grpSparse}
\end{figure}

Under this model, the prior $\xbf$ and the measurements $\ybf$ are naturally
described by a graphical model with linear mixing.
Due to the independence assumptions, the posterior density of $\xbf$ given $\ybf$
factors as
\beq \label{eq:pxySp}
    p(\xbf|\ybf) = \frac{1}{Z(\ybf)} \prod_{i=1}^mp(y_i|z_i)
        \prod_{j=1}^n P(x_j|\xibf_{\gamma(j)})\prod_{k=1}^K P(\xi_k),
\eeq
where $P(x_j|\xibf_{\gamma(j)})$ is the conditional density
for the random variable in \eqref{eq:xxi}.
The factor graph corresponding to this distribution is shown in
Fig.\ \ref{fig:grpSparse}.

Under this graphical model, 
\ifarxiv
Appendix~\ref{sec:grpSparseUpdate} 
\else
\cite[App.~C]{RanganFGBS:17arXiv}
\fi
shows that SP-HyGAMP from Algorithm~\ref{algo:GAMP}
reduces to the simple procedure outlined in Algorithm~\ref{algo:GAMPGrpSparse}.
A similar MS-HyGAMP variant could also be derived.
In lines \ref{line:xhatGrp} and \ref{line:QxGrp} of Algorithm~\ref{algo:GAMPGrpSparse}, we used 
$\Exp(X|R; Q^r, \rhohat)$ and $\var(X|R; Q^r, \rhohat)$ to denote
the expectation and variance, respectively,
of the scalar random variable $X$ with density
\beq \label{eq:xvrho}
    X \sim \left\{ \begin{array}{cl}
        0 & \mbox{with probability } 1-\rhohat \\
        V & \mbox{with probability } \rhohat;
        \end{array} \right.
\eeq
and $R$ is an AWGN corrupted version of $X$,
\beq \label{eq:xrGrp}
    R = X + W, \quad W \sim {\cal N}(0, Q^r).
\eeq

\begin{algorithm}
\caption{SP-HyGAMP for group sparsity}
\begin{algorithmic}[1]  \label{algo:GAMPGrpSparse}
\STATE{ } \COMMENT{Initialization}
\STATE{ $t \gets 0$ }
\STATE{ $Q^r_j(\tm1) \gets \infty$ }
\STATE{ $\LLR_{j \la k}(\tm1) \gets \log(\rho/(1-\rho))$ }
\STATE{ $\rhohat_j(t) \gets 1-\prod_{k \in \gamma(j)}1/(1+\exp \LLR_{j \la k}(\tm1))$ }

\REPEAT
\vspace{0.1in}
\STATE{ }  \COMMENT{Basic GAMP update}
\STATE{ $\xhat_j(t) \gets \Exp(X|R=\rhat_j(\tm1); Q^r_j(\tm1), \rhohat_j(t))$ }
    \label{line:xhatGrp}
\STATE{ $Q^x_j(t) \gets \var(X|R=\rhat_j(\tm1); Q^r_j(\tm1), \rhohat_j(t))$ }
    \label{line:QxGrp}
\STATE{ $\zhat_i(t) \gets \sum_j A_{ij}\xhat_j(t)$ } \label{line:wkBegin}
\STATE{ $Q_i^p(t) \gets \sum_j |A_{ij}|^2Q_j^x(t)$ } \label{line:A_sq}
\STATE{ $\phat_i(t) \gets \zhat_i(t) - Q^p_i(t)\shat_i(\tm1)$ }
\STATE{ $\zhat^0_i(t) \gets \Exp(z_i| \phat_i(t), Q^p_i(t) )$ } \label{line:zhatGrp}
\STATE{ $Q^z_i(t) \gets \var(z_i| \phat_i(t), Q^p_i(t) )$ }\label{line:QzGrp}
\STATE{ $\shat_i(t) \gets (\zhat^0_i - \phat_i(t))/Q^p_i(t)$ }
\STATE{ $Q^s_i(t) \gets Q^{-p}_i(t)(1 - Q^z_i(t)/Q^p_i(t))$ }
\STATE{ $Q^{-r}_j(t) \gets \sum_i |A_{ij}|^2Q^s_i(t)$} \label{line:A_tran_sq}
\STATE{ $\rhat_j(t) \gets \xhat_j(t) + Q^r_j(t)\sum_i A_{ij}\shat_i(t)$} \label{line:gampEnd}

\vspace{0.1in}
\STATE{ } \COMMENT{Sparsity-rate update}
\STATE{ $\rhohat_{j \ra k}(t) \gets 1 - \prod_{i \in \{\gamma(j) \setminus k\}} 1/(1+\exp\LLR_{i \la k}(\tm1))$ }
\STATE{ Compute $\LLR_{j \ra k}(t)$ from \eqref{eq:llrjrk}} \label{line:llrjrk}
\STATE{ $\LLR_{j \la k}(t) \gets \log(\rho/(1-\rho)) + \sum_{i \in \{G_k\setminus j\}} \LLR_{i \ra k}(t)$ }
\STATE{ $\rhohat_j(\tp1) \gets 1-\prod_{k \in \gamma(j)} 1/(1+\exp\LLR_{j \la k}(t))$ }
\STATE{ $t \gets \tp1$ }

\UNTIL{Terminate}

\end{algorithmic}
\end{algorithm}

Algorithm~\ref{algo:GAMPGrpSparse} can be interpreted as the GAMP procedure from \cite{Rangan:11-ISIT}
with an additional update of the sparsity rates.  Specifically,
each iteration $t$ of the algorithm has two stages.
The first stage, labeled as the ``basic GAMP update,"
contains the updates from the basic GAMP algorithm
\cite{Rangan:11-ISIT}, which treats the components $x_j$ as
independent with sparsity rate $\rhohat_j(t)$.
The second stage of Algorithm~\ref{algo:GAMPGrpSparse}, 
labeled as the ``sparsity-rate update," updates the sparsity rates $\rhohat_j(t)$
based on the estimates returned by the first stage.

The second stage of Algorithm \ref{algo:GAMPGrpSparse}
has a simple interpretation.  The quantities $\rhohat_j(t)$ and $\rhohat_{j \ra k}(t)$
can be interpreted, respectively, as estimates for the probabilities
\beqan
    \rho_j &=& \Pr\bigCond{ \xi_k=1 \mbox{ for some } k \in \gamma(j) }{\ybf} \\
    \rho_{j \ra k} &=&
        \Pr\bigCond{\xi_i=1 \mbox{ for some } i \in \{\gamma(j) \setminus k\} }{\ybf}.
\eeqan
That is, $\rhohat_j(t)$ is an estimate of the probability that
the component $x_j$ belongs to at least one
active group and $\rhohat_{j \ra k}(t)$ is an estimate of the probability
that it belongs to an active group other than $G_k$.
Similarly, the quantities $\LLR_{j \ra k}(t)$ and $\LLR_{j \la k}(t)$ are
estimates for the log likelihood ratios
\[
    \LLR_k = \log\frac{P\bigCond{\xi_k=1}{\ybf}}{P\bigCond{\xi_k=0}{\ybf}}.
\]
Most of the updates in the second stage are natural
conversions from LLR values to estimates of $\rho_j$ and $\rho_{j \ra k}$.
In line \ref{line:llrjrk}, the LLR message is computed as
\beq \label{eq:llrjrk}
    \LLR_{j \ra k}(t) = \log\left( \frac{p_R(\rhat_j(t); Q^r_j(t), \rhohat=1)}
        {p_R(\rhat_j(t); Q^r_j(t), \rhohat=\rhohat_{j\ra k}(t))} \right),
\eeq
where $p_R(r; Q^r,\rhohat)$ is the probability density for the
scalar random variable $R$ in \eqref{eq:xrGrp}, where $X$ has the density \eqref{eq:xvrho}.
The message \eqref{eq:llrjrk}
is the ratio of two likelihoods:
the likelihood that $x_j$ belongs to an active group 
and the likelihood that $x_j$ belongs to an active group other than $G_k$.

To summarize, Algorithm~\ref{algo:GAMPGrpSparse} provides a simple and intuitive way to extend the basic GAMP algorithm of \cite{Rangan:11-ISIT} to group-structured sparsity.

The HyGAMP algorithm for group sparsity is also extremely general.
The algorithm can apply to arbitrary priors and output channels.
In particular, the algorithm can incorporate logistic outputs
that are often used for group sparse classification problems
\cite{KimKK:06,MeierVB:08,LozanoSA:11};
details are provided in \cite{Ziniel:TSP:15}.
Also, the method can handle arbitrary, even overlapping, groups.
In contrast, the extensions of other iterative algorithms to the case of overlapping groups
sometimes requires approximations; see, for example, \cite{RaoNowWK:11}.
In fact, the methodology is quite general and likely \text{may} be applied to general structured
sparsity, including possibly the graphical-model-based sparse structures
in image processing considered in \cite{CevICB:10}.

\subsection{Computational Complexity}

In addition to its generality, the HyGAMP procedure is among the most computationally
efficient for group sparsity.  To illustrate this point, consider
the special case when there are $K$ non-overlapping groups of $d$ elements each.
In this case, the total vector dimension for $\xbf$ is $n=Kd$.
We consider the non-overlapping case since there are many algorithms that apply
to this case that we can compare against.
For non-overlapping uniform groups,
Table \ref{tbl:grpSparseComp} compares the computational cost of the HyGAMP
algorithm to other methods.

The computational cost of each iteration of the HyGAMP algorithm, 
Algorithm~\ref{algo:GAMPGrpSparse}, is dominated by the matrix multiplications by $\Abf$
(line~\ref{line:wkBegin}) and $\Abf\tran$ (line~\ref{line:gampEnd})
and by the componentwise squares of $\Abf$ and $\Abf\tran$
(lines~\ref{line:A_sq} and~\ref{line:A_tran_sq}).
Each of these operations has $O(mn) = O(mdK)$ cost.
Note that the multiplications by componentwise-square matrices can be eliminated by using the scalar-variance version of GAMP \cite{Rangan:11-ISIT}.
Also, the multiplications by $\Abf$ and $\Abf\tran$ are relatively cheap if the
matrix has a fast transform (e.g., FFT).
The other per-iteration computations are the $m$ scalar estimates at the output
(lines \ref{line:zhatGrp} and \ref{line:QzGrp}); the
$n$ scalar estimates at the input (lines \ref{line:xhatGrp} and \ref{line:QxGrp});
and the updates of the LLRs.  All of these computations are relatively simple.

For the case of non-overlapping groups, the HyGAMP algorithm could also
be implemented using vector-valued components.  Specifically, the
vector $\xbf$ can be regarded as a block vector with $K$ vector components, each  of dimension $d$.
The general HyGAMP algorithm, Algorithm \ref{algo:GAMP}, can be applied on the vector-valued
components.  To contrast this with Algorithm \ref{algo:GAMPGrpSparse}, we will call
Algorithm \ref{algo:GAMPGrpSparse} HyGAMP with scalar components, and call
the vector-valued case HyGAMP with vector components.

The cost is slightly higher for HyGAMP with vector components.  In this case, there are no
non-trivial strong edges since the block components are independent.  However, in the update \eqref{eq:Qpupdate}, each $\Abf_{ij}$ is $1 \x d$ and $\Qbf^x_j(t)$ is $d \x d$.  Thus,
 the computation \eqref{eq:Qpupdate} requires $mK$ computations of $d^2$ cost each for a total cost
of $O(mKd^2) = O(mnd)$, which is the dominant cost.  Of course, there may be a benefit
in performance for HyGAMP with vector components, since it maintains the complete correlation
matrix of all the components in each group.  We do not investigate this possible performance
benefit in this paper.

Also shown in Table \ref{tbl:grpSparseComp} is the cost of the relaxed BP method from \cite{KimCJBY:11},
which also uses approximate message passing similar to HyGAMP with vector components.
That method, however,  performs the same computations as HyGAMP on each of the $mK$
graph edges as opposed to the $m+K$ graph vertices. It can be verified that the resulting cost
has an $O(mK^2d^2)=O(mn^2)$ term.

\begin{table}
\begin{center}
\begin{tabular}{|p{1.9in}|p{1.1in}|}
\hline
\textbf{Method} & \textbf{Complexity}\\ \hline
Group-OMP \cite{LozanoSA:08} & $O(\rho mn^2)$ \\ \hline
Group-Lasso  \cite{YuanLin:06,ZhaoRY:08,WrightNF:09} & $O(mn)$ per iteration \\ \hline
Relaxed BP with vector components  \cite{KimCJBY:11} & $O(mn^2)$ per iteration \\ \hline
HyGAMP with vector components & $O(mnd)$ per iteration \\ \hline
HyGAMP with scalar components & $O(mn)$ per iteration \\ \hline
\end{tabular}
\end{center}
\caption{Complexity comparison for different algorithms for
group sparsity estimation of a sparse vector with  $K$ groups, each group of dimension $d$.
The number of measurements is $m$ and the sparsity rate is $\rho$.}
\label{tbl:grpSparseComp}
\end{table}


These message passing algorithms can be compared against widely-used group LASSO methods
\cite{YuanLin:06,ZhaoRY:08}, which estimate $\xbf$ by solving some variant of
a regularized least-squares problem of the form
\beq \label{eq:xhatLasso}
    \xbfhat := \argmin_{\xbf} \frac{1}{2}\|\ybf-\Abf\xbf\|^2 + \gamma\sum_{j=1}^n\|\xbf_j\|_2,
\eeq
for some regularization parameter $\gamma > 0$.  The problem \eqref{eq:xhatLasso} is convex
and can be solved via a number of methods including \cite{FigueiredoWN:07,KimKLBG:07,WrightNF:09}, the
fastest of which is the SpaRSA algorithm of \cite{WrightNF:09}.  Interestingly, this algorithm
is similar to the GAMP method in that the algorithm is an iterative
procedure, where in each iteration there is a linear update followed by a componentwise
scalar minimization.  Like the GAMP method, the bulk of the cost is the $O(mn)$
operations per iteration for the linear transform.
An alternative approach for group sparse estimation
is group orthogonal matching pursuit (Group-OMP) of \cite{LozanoSA:08,LozanoSA:11},
a greedy algorithm that
detects one group at a time.  Each round of detection requires $K$ correlations of cost
$md^2$.  If there are on average $\rho K$ nonzero groups, the total complexity will be
$O(\rho K^2md^2)=O(\rho mn^2)$.
From the complexity estimates summarized in Table \ref{tbl:grpSparseComp} it can be seen
that GAMP, despite its generality, is computationally as simple (per iteration)
as some of the most efficient algorithms specifically designed for the group sparsity problem.

Of course, a complete comparison requires that we consider the number of iterations,
not just the computation per iteration.
This comparison requires further study beyond the scope of this paper.
However, it is possible that the HyGAMP procedure will be favorable in this regard.
Our simulations below show good convergence after only 10--20 iterations.
Moreover, in the case of independent (i.e.\ non-group) sparsity,
the number of iterations for AMP algorithms is typically small and often much less than
other iterative methods. 
Examples in~\cite{Montanari:12-bookChap} show excellent convergence
in $10$ to $20$ iterations, which is dramatically faster than the iterative soft-thresholding method
of \cite{DaubechiesDM:04}.

\subsection{Numerical Experiments}

\begin{figure}
\begin{center}
  \psfrag{Normalized MSE (dB)}[b][b][0.8]{\sf Normalized MSE (dB)}
  \psfrag{Num measurements (m)}[t][t][0.8]{\sf Number of Measurements $m$}
  \includegraphics[width=3in,clip]{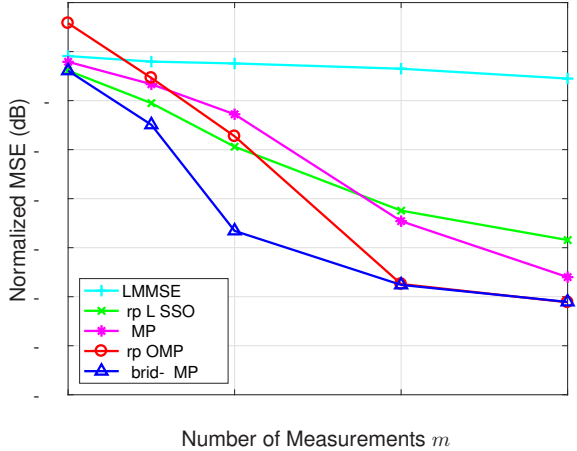}
\end{center}
\caption{Comparison of performances of various estimation algorithms for group
sparsity with $n=100$ groups of dimension $d=4$ with a sparsity fraction of $\rho=0.1$.}
\label{fig:grpSparseSim}
\end{figure}

Fig.~\ref{fig:grpSparseSim} shows a simple simulation comparison of the mean
squared error (MSE) of the HyGAMP method
(Algorithm \ref{algo:GAMPGrpSparse}) along with group OMP, group LASSO, 
\textb{basic GAMP},
and the simple linear MMSE estimator.
The simulation used a vector $\xbf$ with $n=100$ groups of size $d=4$ and sparsity
fraction of $\rho=0.1$.  The matrix was i.i.d.\ Gaussian and the observations were
with AWGN noise at an SNR of 20 dB\@.  The number of measurements $m$ was varied from
50 to 200, and the plot shows the MSE for each of the methods.
The HyGAMP method was run with 20 iterations.
In group LASSO, at each value of $m$, the algorithm was simulated with several values of the
regularization parameter $\gamma$ in \eqref{eq:xhatLasso} and the plot shows the
minimum MSE\@.  In Group-OMP, the algorithm was run with the true value of the number of nonzero
coefficients.  It can be seen that the HyGAMP method is consistently as good or better than
both other methods.
\textb{Furthermore, HyGAMP is significantly better than basic GAMP, which exploits sparsity but not group sparsity.}
All code for the simulations can be found in the GAMPmatlab package 
\cite{Rangan:GAMP-code}.

We conclude that, for the problem of group-sparse recovery from AWGN-corrupted measurements,
the HyGAMP method is at least comparable in performance and computational complexity to the
most competitive algorithms.  On top of this, HyGAMP offers a much more general framework
that can include more rich modeling in both the output and input.


\section{Application to Multinomial Logistic Regression} \label{sec:mlr}

In a second example of the HyGAMP method, we apply it to the problem of \emph{multiclass linear classification} using the approach known as \emph{multinomial logistic regression}.

\subsection{Multinomial Logistic Regression}

In \emph{multiclass classification} \cite{Bishop:06}, one observes a training set $\{(\abf_i,y_i)\}_{i=1}^m$ consisting of $m$ pairs of a feature vector $\abf_i\in\mathbb{R}^n$ and a $d$-ary class label $y_i\in\{1,...,d\}$.
The goal is then to infer the unknown $d$-ary class label $y_0$ of an observed feature vector $\abf_0$. 
In the \emph{linear} approach to this problem, we design a weight matrix $\Xbfhat \in \mathbb{R}^{n \times d}$ from the training set.
Then, given an unlabeled feature vector $\abf_0$, we first generate a vector of linear ``scores'' $\zbf_0 := \Xbfhat\tran \abf_0 \in \mathbb{R}^d$, and estimate the class label $y_0$ as the index of the largest score, i.e.,
\begin{equation}
\hat{y}_0 = \argmax_k [\zbf_0]_k.
\end{equation} 

\newcommand{\mlr}{_{\text{\sf mlr}}}
\newcommand{\lap}{_{\text{\sf lap}}}
\newcommand{\bg}{_{\text{\sf bg}}}
Multinomial linear regression (MLR) \cite{Bishop:06} is one of the best known methods to design the weight matrix $\Xbf$.
There, the labels $\{y_i\}$ are modeled as conditionally independent given the scores $\{\zbf_i\}$, where $\zbf_i := \Xbf\tran \abf_i$.
That is,
\begin{subequations}
\label{eqs:mlr}
\begin{equation}
\text{Pr}(\ybf | \Xbf;\Abf) 
= \prod_{i=1}^m p\mlr(y_i |\Xbf\tran \abf_i), 
\label{eq:like}
\end{equation}
where $p\mlr(y_i |\zbf_i)$ is the multinomial logistic pmf,
\begin{equation}
p\mlr(y_i | \zbf_i) := \frac{\exp\big([\zbf_i]_{y_i}\big)}{\sum_{k=1}^d \exp\big([\zbf_i]_k\big)}, \quad y_i \in \{1,...,d\}.
\label{eq:mlr}
\end{equation}
\end{subequations}
The rows $\xbf_j\tran$ of the weight matrix $\Xbf$ are then modeled as i.i.d., 
\begin{equation}
p(\Xbf) = \prod_{j=1}^n p(\xbf_j).
\end{equation}

For log-convex $p(\xbf_j)$, MAP estimation of $\Xbf$ is a convex problem.
The log-convex Laplacian prior
\begin{equation}
p\lap(\xbf_j) 
= \left(\lambda/2\right)^d \exp\big(-\lambda \|\xbf_j\|_1\big)
\label{eq:laplacian}
\end{equation}
is a popular choice for $p(\xbf_j)$ that promotes sparsity in the designed weight matrix $\Xbfhat$.
Sparsity is essential in the case that the feature dimension $n$ is much larger than the number of training examples $m$. 
Fast implementations of sparse MLR were proposed in \cite{KrishnapuramCFH:05} and refined in \cite{FriedmanHT:10}.

\subsection{HyGAMP Algorithm}

Max-sum HyGAMP (MS-HyGAMP) can be directly applied to solve the above optimization problem.
To do this, we set $\Abf_{ij} = [\abf_i]_j \Ibf_d~\forall i,j$ 
and, recalling \eqref{eq:fi}, we choose 
$f_i(\zbf_i)=\log p\mlr(y_i|\zbf_i)~\forall i=1,...,m$,
and recalling \eqref{eq:fmj}, we choose
$f_{m+j}(\xbf_j)=\log p\lap(\xbf_j)~\forall j=1,...,n$.
Then \eqref{eq:xupdate1} boils down to 
\begin{align}
\xbfhat_j
&= \arg\min_{\xbf} \frac{1}{2}(\xbf-\rbfhat_j)\tran [\Qbf^r_j]^{-1} (\xbf-\rbfhat_j) + \lambda\|\xbf\|_1 ,
\label{eq:xhat_ms}
\end{align}
and \eqref{eq:xz0max} boils down to
\begin{align}
\zbfhat_i
&= \arg\min_{\zbf} \frac{1}{2}(\zbf-\pbfhat_i)\tran [\Qbf^p_i]^{-1} (\zbf-\pbfhat_i) - \log p\mlr(y_i|\zbf) 
\label{eq:zhat_ms} .
\end{align}
Both problems are convex and can be solved using standard methods, e.g., majorization--minimization or Newton's method in the case of \eqref{eq:zhat_ms}.
For more details, 
including the implementation of \eqref{eq:xupdate2} and \eqref{eq:Dz}, 
we refer the reader to \cite{Byrne:15}.

SP-HyGAMP can also be applied to MLR, again using the likelihood
\eqref{eqs:mlr}.
However, rather than the Laplacian prior \eqref{eq:laplacian}, we suggest choosing the Bernoulli-multivariate-Gaussian prior
\begin{subequations} \label{eq:BGvec} 
\begin{align}
p(\Xbf)
&=\prod_{j=1}^n p\bg(\xbf_j) \\
p\bg(\xbf_j) 
&= \beta \delta(\xbf_j) + (1 - \beta) \mathcal{N}(\xbf_j;\zero, q\Ibf)
\end{align}
\end{subequations}
with $\beta\in[0,1)$, which promotes approximate row-sparsity in $\Xbfhat$ under sum-product inference.
In this case, it can be shown \cite{Byrne:15} that \eqref{eq:xupdateSp} can be computed in closed form as
\begin{align}
C_n
&= 1+\frac{1-\beta}{\beta}\frac{\mathcal{N}(\zero;\rbfhat_j,\Qbf^r_j)}
        {\mathcal{N}(\zero;\rbfhat_j,q\Ibf+\Qbf^r_j)} \\
\xbfhat_j
&= \frac{1}{C_n} \left(\Ibf+\frac{1}{q}\Qbf_j^r\right)^{-1} \rbfhat_j \\
\Qbf^x_j
&= \frac{1}{C_n}\left(\Ibf+\frac{1}{q}\Qbf^r_j\right)^{-1}\Qbf^r_j
        + (C_n-1)\xbfhat_j\xbfhat_j\tran .
\end{align}
Although we are not aware of a closed-form solution to \eqref{eq:zDsp}, it can be approximated using numerical integration.

\subsection{Numerical Experiments}

We will now describe the results of two experiments used to evaluate the application of HyGAMP to sparse MLR\@.
In these experiments, SP-HyGAMP and MS-HyGAMP were compared to two state-of-the-art sparse MLR algorithms: SBMLR from \cite{CawleyTG:07} and GLMNET from \cite{FriedmanHT:10}.

\subsubsection{Synthetic Data} 

We first performed an experiment on synthetic data with 
$d=3$ classes, $n=500$ features, and $m=102$ examples.
The use of synthetic data allowed us to analytically compute the expected test-error rate associated with the designed weight matrices $\Xbfhat$. 

To generate the synthetic data, we first constructed the set of
training labels $\{y_i\}$ such that 
$m/d$ training samples were dedicated to each class.
Then we drew feature vectors $\{\abf_i\}$ i.i.d.\ from the class-conditional density $\abf_i | y_i \sim \mathcal{N}(\vec{\mu}_{y_i}, v \Ibf_n)$.
The class means $\{\vec{\mu}_{y}\}_{y = 1}^d$ were $10$-sparse, with support chosen uniformly at random and with non-zero entries chosen uniformly from the columns of a $10\times 10$ random orthonormal matrix.
The parameter $v$ was then chosen to achieve a Bayes error rate of $10\%$.
Thus, only $10$ of the $500$ features were discriminatory.
Note that the data-generation model is not matched to the statistical model assumed \textb{in the derivation of} MS-HyGAMP or SP-HyGAMP\@.

To test the algorithms, we performed $12$ trials, where in each trial we invoked each algorithm-under-test on randomly generated training data and then computed the resulting expected test-error rate.
The SP-HyGAMP algorithm used \eqref{eq:BGvec} with parameters $(\beta,q)$ tuned over a $3\times 5$ logarithmically-spaced grid using 5-fold cross-validation (CV).
The GLMNET algorithm, which solves the same convex optimization problem as MS-HyGAMP, tuned $\lambda$ in \eqref{eq:laplacian} over $25$ logarithmically-spaced values using $5$-fold CV\@.
The same CV-optimal $\lambda$ was then used for MS-HyGAMP\@.
Finally, SBMLR is parameter-free, and thus did not require tuning.

For a designed weight matrix $\Xbfhat=[\xbfhat_1,...,\xbfhat_d]$, 
the expected test-error rate can be analytically computed \cite{Byrne:15} as 
\begin{align}
\Pr\{\text{err}\}
&= 1-\frac{1}{d}\sum_{y=1}^d \Pr\{\text{cor}|y\} 
\label{eq:Perr}\\
\Pr\{\text{cor}|y\}
&= \Pr \bigcap_{k\neq y} \left\{ (\xbfhat_y-\xbfhat_k)\tran \abf < (\xbfhat_y-\xbfhat_k)\tran\vec{\mu}_y \right\} ,
\label{eq:Perry}
\end{align}
where $\abf\sim\mathcal{N}(\zero,v\Ibf_n)$ and the multivariate normal cdf in \eqref{eq:Perry} was computed using Matlab's \texttt{mvncdf}.

In addition to computing the expected test-error rate, we computed two metrics for the sparsity of the designed weight matrices.
The metric $\hat{K}_{\ell_0}=\|\Xbfhat\|_0$ quantifies absolute sparsity, i.e., the number of non-zero elements in $\Xbfhat$.
But since the weights returned by SP-HyGAMP are non-zero with probability one, we also computed the ``effective sparsity'' $\hat{K}_{99}$, which is defined as the minimum number of elements in $\Xbfhat$ required to reach $99\%$ of $\|\Xbfhat\|_F^2$.

Table~\ref{tab:synth_mlr} shows the expected test-error rate, $\hat{K}_{99}$, and $\hat{K}_{\ell_0}$ of each algorithm, averaged over 12 independent trials.
From this table, we see that MS-HyGAMP and GLMNET matched on all metrics.
This result is expected because the two algorithms aim to solve the same convex problem, and it offers evidence that they do in fact solve the problem.
Thus, in the sequel, we report only the results of GLMNET\@.
Next, Table~\ref{tab:synth_mlr} shows that the SP-HyGAMP achieved the best expected test-error rate of $13.981\%$, with SBMLR achieving the second best.
For comparison, we recall that the Bayes (i.e., minimum) expected error rate was $10\%$ in this experiment.
The table also shows that the (average) effective sparsity $\hat{K}_{99}$ was similar for all algorithms, and smaller than the sparsity of the Bayes' optimal classifier for this dataset, which is $\hat{K}_{\ell_0} = 30$. 

\begin{table}
\begin{center}
\begin{tabular}{|l|c|c|c|}
\hline
Algorithm   &  \% Error & $\hat{K}_{99}$  &  $\hat{K}_{\ell_0}$\rule{0mm}{3mm} \\ \hline
GLMNET      & 14.787  & 13.25  & 25.75 \\ 
MS-HyGAMP   & 14.787  & 13.25  & 25.75 \\
SBMLR       & 14.059  & 15.08  & 28.92 \\ 
SP-HyGAMP   & \bf 13.981  & 16.08 & 1500    \\
\hline
\end{tabular}
\end{center}
\vspace{2mm}
\caption{Results for the synthetic data experiment}
\label{tab:synth_mlr}
\end{table}

\subsubsection{Handwritten Digit Classification} 

In the second experiment, we tested SP-HyGAMP, GLMNET, and SBMLR on the Mixed National Institute of Standards and Technology (MNIST) dataset\cite{LeCunBBH:98}. 
The MNIST dataset consists of $m=70\,000$ total images of handwritten digits $0$ through $9$, hence $d=10$.
Each image has $n=784$ pixels.
In this experiment we performed 24 trials, where in each trial we randomly partitioned the total dataset into a training and testing portion.
Within each trial, we varied the number of image samples in the training partition from $m=56$ to $m=1000$.
Using the training data, we used each algorithm-under-test to design a weight matrix, which was then used to compute an empirical error-rate on the test partition of the dataset.
In this experiment, SP-HyGAMP and GLMNET tuned their associated parameters in a similar manner as in the synthetic experiment. 
However, they used $2$-fold CV instead of $5$-fold CV to reduce computation.

\begin{figure}
\begin{center}
\newcommand{\sz}{0.8}
\psfrag{test error rate}[b][b][\sz]{\textsf{Test-Error Rate}}
\psfrag{training samples}[t][t][\sz]{\textsf{Number of Training Samples $m$}}
\includegraphics[width=3.3in]{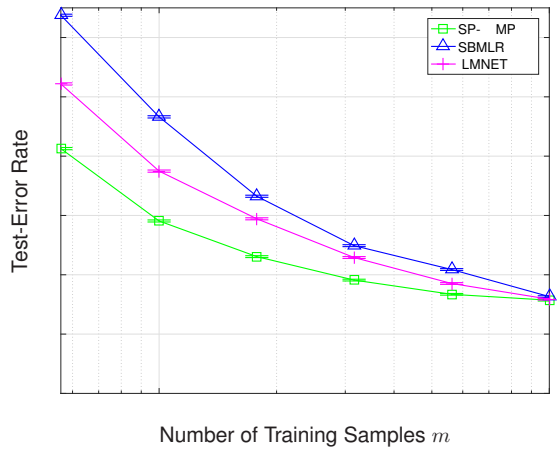}
\end{center}
\caption{Classification results for MNIST dataset.
}
\label{fig:mnist}
\end{figure}

Figure~\ref{fig:mnist} shows the empirical test-error rate versus the number of training samples $m$, averaged over the $24$ random trials.
The error bars indicate the standard deviation of the empirical error-rate estimate.
The figure shows that, for all $m$, SP-HyGAMP achieved the best test-error rate and GLMNET achieved the second best. 
The figure also shows that, for all algorithms, the test-error rate decreased 
\textb{to a common value as the number of training samples $m$ increased. 
This is not surprising; we expect that, with enough training data, any reasonable approach should recover a close approximation to the Bayes-optimal linear classifier.
A much more difficult problem is designing a good linear classifier from limited training data, and, for this problem, Figure~\ref{fig:mnist} shows that SP-HyGAMP beats the competition.}

%
%
%


\blue
\subsection{Simplified HyGAMP and EM/SURE Tuning} \label{sec:SHyGAMP}

When directly applied to multinomial logistic regression, each iteration of HyGAMP involves the update of $O(m+n)$ multivariate Gaussian pdfs, each of dimension $d$, for a total complexity of $O((m+n)d^3)$ per iteration.
This complexity can be quite large in practice, especially relative to state-of-the-art methods like GLMNET and SBMLR\@.
Furthermore, in its more direct form, HyGAMP assumes knowledge of the statistical parameters of its prior and likelihood.
In order to tune these parameters to the data, it was suggested above to use cross-validation (as with GLMNET). 
But $K$-fold cross-validation of $P$ parameters using $G$ hypothesized values of each parameter requires the training and evaluation of $KG^P$ classifiers, which can be very expensive in practice.

Fortunately, for multinomial logistic regression, it is possible to modify HyGAMP in such a way that the complexity of the resulting method becomes competitive with GLMNET and SBMLR\@.
The modification consists of two parts: 
i) a \emph{simplification} of HyGAMP wherein the covariance matrices $\Qbf^r_j,\Qbf^x_j,\Qbf^p_i,\Qbf^z_i$ are constrained to be diagonal; and
ii) an application of EM-based \cite{vila2013expectation} and SURE-based \cite{mousavi2013sure} parameter tuning to the priors and likelihoods relevant to multinomial logistic regression.
A complete description of EM/SURE-tuned simplified HyGAMP (SHyGAMP) for multinomial logistic regression can be found in \cite{ByrneS:16}, with full derivations in \cite{Byrne:15}.
In \cite{ByrneS:16}, a detailed numerical study establishes that EM/SURE-tuned SHyGAMP is competitive in both performance and complexity with GLMNET and SBMLR\@.
Due to space limitations, we refer the interested reader to \cite{Byrne:15} and \cite{ByrneS:16} for more details.

We conclude by saying that, although the ``direct'' application of HyGAMP from Section~\ref{sec:gampAlgo} may not lead to a complexity that is always competitive with state-of-the-art methods, it acts as an important \emph{first step} in deriving simplified and/or enhanced version of HyGAMP\@.
This underscores the importance of HyGAMP as stated in Section~\ref{sec:gampAlgo}.

\black
\section{Conclusions}
A general model for optimization and statistical inference
based on graphical models with linear mixing was presented.  The linear mixing components
of the graphical model account for interactions through aggregates of large numbers of
small, linearizable perturbations.  Gaussian and second-order approximations are shown
to greatly simplify the implementation of loopy BP for these interactions,
and the HyGAMP framework presented here enables these approximations to be incorporated in
a systematic manner in general graphical models.  
Simulations
were presented for group sparsity and multinomial logistic regression, where the HyGAMP method has equal or superior performance to existing methods. 
\textb{Although we saw that, in multinomial logistic regression, a direct application of HyGAMP does not lead to state-of-the-art computationally complexity, a modification of the HyGAMP presented here suffices to address the complexity issue \cite{Byrne:15,ByrneS:16}.}
The generality of the proposed HyGAMP algorithm also allows its application to many other problems beyond these two examples, 
\blue
such as
multiuser detection in massive MIMO \cite{wang2015mimo,wang2016signal},
inference for neuronal connectivity \cite{fletcher2014scalable},
fitting neural mass spatio-temporal models \cite{fletcher2015neural},
user activity detection in cloud-radio random access \cite{utkovski2017random},
and
decoding from pooled data \cite{alaoui2017pooled}.
\black
In addition to pursuing such applications, future work will focus on
establishing rigorous theoretical analyses along the lines of
\cite{BayatiM:11,Rangan:11-ISIT} for specific instances of HyGAMP.

\appendices

\ifarxiv
  \section{Derivation of SP-HyGAMP} \label{sec:SumProdDeriv}

\subsection{Preliminary Lemma}
Before deriving the SP-HyGAMP algorithm, we need the following result.
Let $H(\wbf,\vbf)$ be a real-valued function of vectors $\wbf$ and $\vbf$ of the form
\beq \label{eq:Huv}
    H(\wbf,\vbf) = H_0(\wbf) - \frac{1}{2}\|\wbf-\vbf\|^2_{\Qbf^v}
\eeq
for some positive definite matrix $\Qbf^v$.

\begin{lemma}  \label{lem:logzDeriv}
Suppose that $\Wbf$ and $\Vbf$ are random vectors with a
conditional probability distribution function of the form
\[
    p_{\Wbf|\Vbf}(\wbf \MID \vbf) = \frac{1}{Z(\vbf)}\exp\left[u H(\wbf,\vbf)\right],
\]
where $H(\wbf,\vbf)$ is given in \eqref{eq:Huv}, $u > 0$ is some constant and
$Z(\vbf)$ is a normalization constant (called the partition function).
Then,
\begin{subequations} \label{eq:logZRel}
\beqa
     \frac{\partial}{\partial \vbf}\xbfhat(\vbf)
         &=& \Dbf\Qbf^{-v}  \label{eq:expDeriv} \\
     \frac{\partial}{\partial \vbf}\log Z(\vbf) &=&
        \Qbf^{-v}(\xbfhat(\vbf)-\vbf)  \label{eq:logZD1} \\
     \frac{\partial^2}{\partial \vbf^2}\log Z(\vbf) &=& -\Qbf^{-v}
     + \Qbf^{-v}\Dbf\Qbf^{-v} \label{eq:logZD2}
\eeqa
where
\[
   \xbfhat(\vbf) =  \Exp[\Wbf \MID \Vbf=\vbf], \ \ \Dbf = u\cov(\Wbf \MID \Vbf=\vbf).
\]
\end{subequations}
\end{lemma}
\begin{IEEEproof}  The relations are standard properties of exponential families
\cite{WainwrightJ:08}.
\end{IEEEproof}

\subsection{SP-HyGAMP Approximation}

First partition the objective function $H_{i \ra j}(\cdot)$
in \eqref{eq:HijBp} as
\beqa
    \lefteqn{ H_{i \ra j}(t,\xbf_{\partial(i)},\zbf_i) } \nonumber \\
    &=& H^{\rm strong}_{i \ra j}(t,\xbf_{\alpha(i)},\zbf_i)
    + H^{\rm weak}_{i \ra j}(t,\xbf_{\beta(i)}),\label{eq:HijPart}
\eeqa
where
\begin{subequations} \label{eq:HijStrWk}
\beqa
    \lefteqn{ H^{\rm strong}_{i \ra j}(t,\xbf_{\alpha(i)},\zbf_i)} \nonumber \\
    & := & f_i(\xbf_{\alpha(i)}, \zbf_i)
               + \sum_{r \in \{ \alpha(i) \setminus j\}} \Delta_{i \la r}(t,\xbf_r),
        \label{eq:HijStr} \\
    \lefteqn{ H^{\rm weak}_{i \ra j}(t,\xbf_{\beta(i)}) :=
        \sum_{r \in \{ \beta(i) \setminus j\}} \Delta_{i \la r}(t,\xbf_r). }
        \label{eq:HijWk}
\eeqa
\end{subequations}
That is, we have separated the terms in $H_{i \ra j}(\cdot)$
between the strong and weak edges.

Then, the marginal distribution $p_{i \ra j}(t,\xbf_j)$ of the distribution
$p_{i \ra j}(t,\xbf_{\partial(i)})$ in \eqref{eq:pijFnBpSp}
can be re-written as
\beqa
    \lefteqn{ p_{i \ra j}(t,\xbf_j) = \int
        p_{i \ra j}(t,\xbf_{\partial(i)})d\xbf_{\partial(i) \backslash j} } \nonumber \\
    &\propto& \int
    \psi^{\rm strong}_{i \ra j}(t,\xbf_j,\zbf_i)
    \psi^{\rm weak}_{i \ra j}(t,\xbf_j,\zbf_i)d\zbf_i, \label{eq:pijSpPart}
\eeqa
where
\begin{subequations}
\beqa
    \lefteqn{ \psi^{\rm strong}_{i \ra j}(t,\xbf_j, \zbf_i) } \nonumber \\
        &\propto& \int\limits_{\xbf_{\alpha(i) \backslash j}}
        \exp\left[uH^{\rm strong}_{i \ra j}(t,\xbf_{\alpha(i)},\zbf_i)\right]
        d\xbf_{\alpha(i) \backslash j}
        \label{eq:DelFnStrSpij} \\
    \lefteqn{ \psi^{\rm weak}_{i \ra j}(t,\xbf_j, \zbf_i) } \nonumber \\
        &\propto& \int\limits_{\substack{\xbf_{\beta(i) \backslash j} \\
        \zbf_i=\Abf_i\xbf} }
        \exp\left[u H^{\rm weak}_{i \ra j}(t,\xbf_{\beta(i)})\right]
            d\xbf_{\beta(i) \backslash j}
        \label{eq:DelFnWkSpij}
\eeqa
\end{subequations}
and the integration in \eqref{eq:DelFnStrSpij} is over the variables
$\xbf_r$ with $r \in \alpha(i) \setminus j$, and
and the integration in \eqref{eq:DelFnWkSpij} is over the variables
$\xbf_r$ with $r \in \beta(i) \setminus j$, and $\zbf_i = \Abf_i\xbf$.

To approximate $p_{i \ra j}(t,\xbf_j)$ in \eqref{eq:pijSpPart},
we separately consider the cases when $(i,j)$ is weak edge
and when it is a strong edge.
We begin with the weak edge case.  That is, $j \in \beta(i)$.
Let
\begin{subequations} \label{eq:xQxDelSp}
\beqa
    \xbfhat_{j}(t) &:=& \Exp[ \xbf_j \MIDD \Delta_{j}(t,\cdot) ],
        \label{eq:xhatDelj} \\
    \xbfhat_{i \la j}(t) &:=& \Exp[ \xbf_{j} \MIDD \Delta_{i \la j}(t,\cdot) ],
        \label{eq:xhatDelij} \\
    \Qbf^x_{j}(t) &:=& u\,\cov[ \xbf_j \MIDD \Delta_{j}(t,\cdot) ]
        \label{eq:QxDelj}  \\
    \Qbf^x_{i \la j}(t) &:=& u\,\cov[ \xbf_j \MIDD \Delta_{i \la j}(t,\cdot) ],
        \label{eq:QxDelij}
\eeqa
\end{subequations}
where
we have used the notation $\Exp[g(\xbf) \MIDD \Delta(\cdot)]$ from \eqref{eq:expDel}.

Now, using the expression for $H^{\rm weak}_{i \ra j}(t,\xbf_{\beta(i)})$
in \eqref{eq:HijWk}, it can be verified that
$\psi^{\rm weak}_{i \ra j}(t,\xbf_j, \zbf_i)$ is equivalent to the
probability distribution of a random variable
\beq \label{eq:ziSpPf}
    \zbf_i = \Abf_{ij}\xbf_j + \sum_{r \in \{\beta(i)\setminus j\}}\Abf_{ir}\xbf_r,
\eeq
with the variables $\xbf_r$ being independent with probability distribution
\[
    p(\xbf_r) \propto \exp(u\Delta_{i \la r}(\xbf_r)).
\]
Moreover, $\xbfhat_{i \la j}(t)$ and $\Qbf^x_{i \la j}(t)/u$ in \eqref{eq:xQxDelSp}
are precisely the mean and variance of the random variables $\xbf_j$
under this distribution.
Therefore, if the summation in \eqref{eq:ziSpPf} is over a large number of terms,
we can then use the CLT to
approximate the variable in $\zbf_i$ in \eqref{eq:ziSpPf} as Gaussian,
with distribution $\psi^{\rm weak}_{i \ra j}(t,\xbf_j, \zbf_i)$ given by
\beq \label{eq:zijSpGauss}
    \psi^{\rm weak}_{i \ra j}(t,\xbf_j, \zbf_i) \approx
    {\cal N}(\Abf_{ij}\xbf_j + \pbfhat_{i \ra j}(t), \Qbf^p_{i \ra j}(t)/u),
\eeq
where 
\begin{subequations} \label{eq:pQpij}
\beqa
    \pbfhat_{i \ra j}(t) &=& \sum_{r \in \{\beta(i) \setminus j\}}
        \Abf_{ir}\xbfhat_{i \la r}(t) \\
    \Qbf^p_{i \ra j}(t) &=& \sum_{r \in \{\beta(i) \setminus j\}}
        \Abf_{ir}\Qbf^x_r(t)\Abf_{ir}^*.
\eeqa
\end{subequations}
Substituting this Gaussian approximation into the probability distribution
$p_{i \ra j}(t,\xbf_{\partial(i)},\zbf_i)$ in \eqref{eq:pijFnBpSp},
and then using the definitions in \eqref{eq:HijStr} and
\eqref{eq:DelFnStrSpij}, we obtain the following approximation
of the message in \eqref{eq:DelijFnBpSp},
\beq \label{eq:DelijG2}
    \Delta_{i \ra j}(t,\xbf_j)\approx G_{i}(t,\Abf_{ij}\xbf_j +
    \pbfhat_{i \ra j}(t), \Qbf^p_{i \ra j}(t)) ,
\eeq
where
\beqa
    \lefteqn{ G_{i}(t,\pbf_i,\Qbf^p_i)  } \nonumber \\
    &:=&  \frac{1}{u}\log
    \int  \exp\left[ uH^z_{i}(t,\xbf_{\alpha(i)},\zbf_i,\pbf_i,\Qbf^p_i) \right]
    d\xbf_{\alpha(i)} d\zbf_i \hspace{0.2in}  \label{eq:GiSp}
\eeqa
and where $H^z_i(\cdot)$ is given in \eqref{eq:Hquadi}.

Now define
\begin{subequations} \label{eq:pQpi}
\beqa
    \pbfhat_{i}(t) &=& \sum_{r \in \beta(i)}
        \Abf_{ir}\xbfhat_{i \la r}(t) \\
    \Qbf^p_{i}(t) &=& \sum_{r \in \beta(i)}
        \Abf_{ir}\Qbf^x_r(t)\Abf_{ir}^*,
\eeqa
\end{subequations}
so that the expressions in \eqref{eq:pQpij} can be re-written as
\begin{subequations} \label{eq:pQpiDiff}
\beqa
    \pbfhat_{i \ra j}(t) &=& \pbfhat_i(t) - \Abf_{ij}\xbfhat_{i \la j}(t) \\
    \Qbf^p_{i \ra j}(t) &=& \Qbf^p_{i}(t) -
        \Abf_{ij}\Qbf^x_j(t)\Abf_{ij}^*.
\eeqa
\end{subequations}
Also, let
\begin{subequations} \label{eq:supdateSp}
\beqa
    \sbfhat_i(t) &=& \frac{\partial}{\partial \pbfhat}
        G_i(t,\pbfhat_i(t),\Qbf^p_i(t)) \\
    \Qbf^{-s}_i(t) &=& -\frac{\partial^2}{\partial \pbfhat^2}
         G_i(t,\pbfhat_i(t),\Qbf^p_i(t)).
\eeqa
\end{subequations}
Using Lemma \ref{lem:logzDeriv}, one can show that the definitions in
\eqref{eq:supdateSp} agree with the updates \eqref{eq:supdate} where
$\zbfhat^0_i(t)$ and $\Qbf^z_i(t)$
are the mean and covariance of the random variable $\zbf_i$
with the distribution \eqref{eq:piFnSp}.

Applying \eqref{eq:supdateSp}, we can take
a second-order approximation of \eqref{eq:DelijG2} as
\beqa
    \lefteqn{ \Delta_{i \ra j}(t,\xbf_j) \approx \mbox{const} }
    \nonumber \\
    &+&
    \sbfhat_i(t)^*\Abf_{ij}(\xbf_j-\xbfhat_j(t)) - \frac{1}{2}
    \|\Abf_{ij}(\xbf_j-\xbfhat_j(t))\|^2_{\Qbf^s_{i}(t)} \nonumber \\
    &=& \mbox{const} + \left[\Abf_{ij}^*\sbf_i(t)
        +\Abf_{ij}^*\Qbf^s_i(t)\Abf_{ij}\xbfhat_j(t)\right]^*
        \xbf_j \nonumber \\
    & & + \frac{1}{2}\xbf_j^*\Abf_{ij}^*\Qbf^s_i(t)\Abf_{ij}\xbf_j
    \label{eq:DelijQuadWk}
\eeqa
for all weak edges $(i,j)$.

Next consider the case when $j \not \in \beta(i)$ so that $(i,j)$ is a strong edge.
In this case, $\psi^{\rm weak}_{i \ra j}(t,\xbf_j, \zbf_i)$ does not depend
on $\xbf_j$ and a similar calculation as above shows that
\beq \label{eq:ziSpGauss}
    \psi^{\rm weak}_{i \ra j}(t,\xbf_j, \zbf_i) \approx
    \psi^{\rm weak}_{i}(t,\zbf_i) :=
    {\cal N}(\pbfhat_{i}(t), \Qbf^p_{i}(t)/u),
\eeq
where $\pbfhat_i(t)$ and $\Qbf^p_i(t)$ are defined in \eqref{eq:pQpi}.
Substituting the Gaussian approximation \eqref{eq:ziSpGauss} into
\eqref{eq:pijFnBpSp},
and then using the definitions in \eqref{eq:HijStr} and
\eqref{eq:DelFnStrSpij}, one can show that the marginal distribution
$p_{i \ra j}(t,\xbf_j)$ in \eqref{eq:pijFnBpSp} is equal to the marginal
distribution of $p_{i \ra j}(t,\xbf_{\alpha(i)}, \zbf_i)$ in \eqref{eq:pijFnSp}.
Therefore, the message $\Delta_{i \ra j}(t,\xbf_j)$
in \eqref{eq:DelijFnBpSp} can be written as \eqref{eq:DelijFnSp}
for all strong edges $(i,j)$.

We now turn to the variable node update \eqref{eq:DelijVarBp} which we
partition as
\beq \label{eq:DelVarijPart}
    \Delta_{i \la j}(t,\xbf_j) = \Delta^{\rm weak}_{i \la j}(t,\xbf_j) +
    \Delta^{\rm strong}_{i \la j}(t,\xbf_j),
\eeq
where
\begin{subequations} \label{eq:DelVarijStrWk}
\beqa
    \Delta^{\rm strong}_{i \la j}(\tp1,\xbf_j) &=&
        \sum_{\ell \neq i ~:~ j \in \alpha(\ell)}
        \Delta_{\ell \ra j}(t,\xbf_j) \label{eq:DelVarijStr} \\
    \Delta^{\rm weak}_{i \la j}(\tp1,\xbf_j) &=&
        \sum_{\ell \neq i ~:~ j \in \beta(\ell)}
        \Delta_{\ell \ra j}(t,\xbf_j). \label{eq:DelVarijWk}
\eeqa
\end{subequations}
Substituting the approximation \eqref{eq:DelijQuadWk} into
\eqref{eq:DelVarijWk} gives
\beq \label{eq:DelVarijQuad}
    \Delta^{\rm weak}_{i \la j}(\tp1,\xbf_j) \approx
     -\frac{1}{2}\|\rbfhat_{i \la j}(t)-\xbf_j\|^2_{\Qbf^r_{i \la j}(t)},
\eeq
where
\begin{subequations} \label{eq:rupdateij}
\beqa
    \Qbf^{-r}_{i \la j}(t) &=& \sum_{\ell \neq i}
        \Abf_{\ell j}^*\Qbf^s_\ell(t)\Abf_{\ell j} \\
    \rbfhat_{i \la j}(t) &=&   \Qbf^{r}_{i \la j}(t) \nonumber \\
        &\times& \left[ \sum_{\ell \neq i} \Abf_{\ell j}^*\sbfhat_\ell(t)
            + \Abf_{\ell j}^*\Qbf^s_\ell(t)\Abf_{\ell j}\xbfhat_j(t) \right]
            \nonumber\\
    &=& \xbfhat(t) + \Qbf^{r}_{i \la j}(t)
        \sum_{\ell \neq i} \Abf_{\ell j}^*\sbfhat_\ell(t).
\eeqa
\end{subequations}
We again consider the case of a weak edge separately from a 
strong edge.  When $(i,j)$ is weak edge,
$j \not \in \alpha(i)$, so that $\Delta^{\rm strong}_{i \la j}(\tp1,\xbf_j)$
in \eqref{eq:DelVarijStr} does not depend on $i$.
Combining \eqref{eq:DelVarijPart} and \eqref{eq:DelVarijQuad},
we see that
\beq \label{eq:DelVarijH1}
    \Delta_{i \la j}(\tp1,\xbf_j) \approx
        H^x_j(t,\xbf_j,\rbfhat_{i \la j}(t),\Qbf^r_{i \la j}(t)),
\eeq
where $H^x_j(\cdot)$ is defined in \eqref{eq:Hxj}.
Also, comparing \eqref{eq:rupdate} with \eqref{eq:rupdateij}, we have
that
\begin{subequations} \label{eq:rupdateij2}
\beqa
    \Qbf^{-r}_{i \la j}(t) &\approx& \Qbf^{-r}_{j}(t)\\
    \rbfhat_{i \la j}(t) &\approx&   \rbfhat_j(t)
        - \Qbf^{r}_{j}(t)\Abf_{i j}^*\sbfhat_i(t).
\eeqa
\end{subequations}
Substituting \eqref{eq:rupdateij2} into  \eqref{eq:DelVarijH1}
we get
\beqa
\lefteqn{
    \Delta_{i \la j}(\tp1,\xbf_j) } \nonumber \\
    &\approx&
        H^x_j(t,\xbf_j,\rbfhat_{j}(t)-\Qbf^{r}_{j}(t)\Abf_{i j}^*\sbfhat_i(t),
    \Qbf^r_{j}(t)). \label{eq:DelVarijH2}
\eeqa
A similar set of calculations shows that $\Delta_j(\tp1,\xbf_j)$
in \eqref{eq:DeljVarBp} can be approximated as
\beq \label{eq:DelVarjH2}
    \Delta_{j}(\tp1,\xbf_j) \approx
        H^x_j(t,\xbf_j,\rbfhat_{j}(t),
    \Qbf^r_{j}(t)).
\eeq
Thus, the definitions of $\xbfhat_j(\tp1)$ and $\Qbf^x_j(\tp1)$ in
 \eqref{eq:xQxDelSp} agree with \eqref{eq:xupdateSp}.

Finally, define
\beq \label{eq:gamDefSp}
    \Gamma_j(t,\rbfhat_j) :=
        \Exp\left[ \xbf_j \MIDD H_j^x(t,\cdot,\rbfhat_j, \Qbf^r_j(\tm1)) \right],
\eeq
where again we are using the notation \eqref{eq:expDel} and $H_j^x(\cdot)$ is
defined in \eqref{eq:Hxj}.
It follows from \eqref{eq:DelVarijH2}, \eqref{eq:DelVarjH2} and
\eqref{eq:xQxDelSp} that
\beqa
    \lefteqn{ \xbfhat_{j}(\tp1)  \approx  \Gamma_j(t,\rbfhat_j(t)) }
        \nonumber \\
    \lefteqn{ \xbfhat_{i \la j}(\tp1) \approx  \Gamma_j(t,\rbfhat_j(t)
        -\Qbf^{r}_{j}(t)\Abf_{i j}^*\sbfhat_i(t)) } \nonumber \\
        &\approx& \xbfhat_j(t)-
    \frac{\partial \Gamma_j(t,\rbfhat_j(t))}{\partial \rbfhat_j}
        \Qbf^{r}_{j}(t)\Abf_{i j}^*\sbfhat_i(t). \label{eq:xhatijGam1Sp}
\eeqa
From the definition \eqref{eq:gamDefSp}, Lemma \ref{lem:logzDeriv} shows that
\beq
    \frac{\partial \Gamma_j(t,\rbfhat_j(t))}{\partial \rbfhat_j}
    \approx \Qbf^{x}(t)\Qbf^{-r}(t),
\eeq
and hence, from \eqref{eq:xhatijGam1Sp},
\beq \label{eq:xhatijGam2Sp}
    \xbfhat_{i \la j}(\tp1) \approx
    \xbfhat_j(\tp1)-
        \Qbf^x(\tp1)\Abf_{i j}^*\sbfhat_i(t).
\eeq
Substituting \eqref{eq:xhatijGam2Sp} into \eqref{eq:pQpi} we obtain
\beqa
   \lefteqn{ \pbfhat_{i}(t) \approx \sum_{j \in \beta(i)}
        \Abf_{ij}\xbfhat_{j}(t) - \sum_{j \in \beta(i)}
        \Abf_{ij}\Qbf^x(t)\Abf_{ij}^*\sbfhat_i(\tm1) } \nonumber \\
    &\approx& \zbf_i(t) - \Qbf^p(t)\sbfhat_i(\tm1),
    \hspace{1.4in} \nonumber
\eeqa
which agrees with the definition in \eqref{eq:pupdate}.

  \section{Derivation of MS-HyGAMP} \label{sec:MaxSumDeriv}

The derivation of MS-HyGAMP is similar to
the derivation of SP-HyGAMP in Appendix~\ref{sec:SumProdDeriv}.

\subsection{Preliminary Lemma}
We begin by stating the analogue to Lemma \ref{lem:logzDeriv}.
For each $\vbf$, let
\begin{subequations} \label{eq:GdefLem}
\beqa
    \wbfhat(\vbf) &:=& \argmax_\wbf H(\wbf,\vbf), \\
    G(\vbf) &:=&  H(\wbfhat(\vbf),\vbf) = \max_\wbf H(\wbf,\vbf),
\eeqa
\end{subequations}
where $H(\wbf,\vbf)$ was given in \eqref{eq:Huv}.

\begin{lemma} \label{lem:GDeriv}  Assume the
maximization in \eqref{eq:GdefLem}
exists and is unique and twice differentiable.
Then,
\begin{subequations} \label{eq:GDeriv}
\beqa
       \frac{\partial}{\partial \vbf} G(\vbf) &=&
        \Qbf^{-v}(\wbfhat(\vbf)-\vbf),
        \label{eq:GDeriv1} \\
    \frac{\partial \wbfhat}{\partial \vbf} &=&  -\Dbf^{-1}\Qbf^{-v},
        \label{eq:uhatDeriv} \\
    \frac{\partial^2}{\partial \vbf^2} G(\vbf) &=&
        -\Qbf^{-v} - \Qbf^{-v}\Dbf^{-1}\Qbf^{-v},      \label{eq:GDeriv2}
\eeqa
\end{subequations}
where
\[
    \Dbf =  \left. \frac{\partial^2 H(\wbf,\vbf)}{\partial \wbf^2}
        \right|_{\wbf=\wbfhat(\vbf)}.
\]
\end{lemma}
\begin{IEEEproof}
Since $\wbf = \wbfhat(\vbf)$ is a maximizer of $H(\wbf,\vbf)$,
\beq \label{eq:uderivLem}
    \frac{\partial H(\wbfhat(\vbf),\vbf)}{\partial \wbf} = 0.
\eeq
Therefore, \eqref{eq:GDeriv1} follows from
\beqan
    \frac{\partial G(\vbf)}{\partial \vbf} & = &
    \frac{\partial H(\wbfhat(\vbf),\vbf)}{\partial \wbf
    \frac{\partial \wbfhat(\vbf)}{\partial \vbf} +
    \frac{\partial H(\wbfhat(\vbf),\vbf) }{\partial \vbf} } \nonumber \\
    &=&    \frac{\partial H(\wbfhat(\vbf),\vbf) }{\partial \vbf}
    = \Qbf^{-v}(\wbfhat(\vbf)-\vbf), \hspace{0.5in}
\eeqan
where the last step is a result of the form of $H(\cdot)$ in \eqref{eq:Huv}.
The form of $H(\cdot)$ in \eqref{eq:Huv} also shows that for all $\wbf$ and $\vbf$
\[
    \frac{\partial^2 H(\wbf,\vbf)}{\partial \wbf\partial \vbf} = \Qbf^{-v}.
\]
Taking the derivative of \eqref{eq:uderivLem},
\[
    \frac{\partial^2 H(\wbfhat,\vbf)}{\partial \wbf\partial \vbf} +
        \frac{\partial^2 H(\wbfhat,\vbf)}{\partial \wbf^2}
        \frac{\partial \wbfhat(\vbf)}{\partial \vbf}  = 0,
\]
which implies that
\[
    \frac{\partial \wbfhat(\vbf)}{\partial \vbf} = -\Dbf^{-1}\Qbf^{-v},
\]
which proves \eqref{eq:uhatDeriv}.
Finally, taking the second derivative of \eqref{eq:GDeriv1}
along with \eqref{eq:uhatDeriv} shows \eqref{eq:GDeriv2}.
\end{IEEEproof}

\subsection{MS-HyGAMP Approximation}

Similar to the SPA derivation,
we first partition the function $H_{i \ra j}(\cdot)$
in \eqref{eq:HijBp} as in \eqref{eq:HijPart}.
We can also partition the maximization \eqref{eq:DelijFnBp} as
\beqa
    \lefteqn{ \Delta_{i \ra j}(t,\xbf_j) } \nonumber \\
    &=& \max_{\zbf_i} \left[
    \Delta^{\rm strong}_{i \ra j}(t,\xbf_j, \zbf_i) +
    \Delta^{\rm weak}_{i \ra j}(t,\xbf_j,\zbf_i) \right],
    \label{eq:bpMaxFnPart}
\eeqa
where
\begin{subequations}
\beqa
    \Delta^{\rm strong}_{i \ra j}(t,\xbf_j, \zbf_i)
        &:=& \max_{\xbf_{\alpha(i) \backslash j}}
        H^{\rm strong}_{i \ra j}(t,\xbf_{\alpha(i)},\zbf_i), \qquad
        \label{eq:DelFnStrij} \\
    \Delta^{\rm weak}_{i \ra j}(t,\zbf_i,\xbf_j)
        &:=& \max_{\substack{\xbf_{\beta(i) \backslash j} \\
        \zbf_i=\Abf_i\xbf} } H^{\rm weak}_{i \ra j}(t,\xbf_{\beta(i)}),
        \label{eq:DelFnWkij}
\eeqa
\end{subequations}
with the maximization in \eqref{eq:DelFnStrij} being over all
$\xbf_r$ with $r \in \alpha(i) \setminus j$;
and the maximization in \eqref{eq:DelFnWkij} over all
$\xbf_r$ with $r \in \beta(i) \setminus j$ subject to $\zbf_i = \Abf_i\xbf$.
The partitioning \eqref{eq:bpMaxFnPart} is valid since the strong
and weak edges are distinct.  This insures that for all $r \in \delta(i)$,
either $r \in \alpha(i)$ or $r \in \beta(i)$, but not both.

The HyGAMP approximation applies to the weak term \eqref{eq:DelFnWkij}.
For any $j$ and all weak edges $(i,j)$, define:
\begin{subequations} \label{eq:xhatQxFn}
\beqa
    \xbfhat_{j}(t) &:=& \argmax_{\xbf_j} \Delta_{j}(t,\xbf_j),
        \label{eq:xhatMaxDelj} \\
    \xbfhat_{i \la j}(t) &:=& \argmax_{\xbf_j} \Delta_{i \la j}(t,\xbf_j),
        \label{eq:xhatMaxDel} \\
    \Qbf^{-x}_{j}(t) &:=& -\frac{\partial^2}{\partial \xbf_j^2}
        \left. \Delta_{j}(t,\xbf_j) \right|_{\xbf_j = \xbfhat_{j}(t)},
        \label{eq:QxMaxDelj}  \\
    \Qbf^{-x}_{i \la j}(t) &:=& -\frac{\partial^2}{\partial \xbf_j^2}
        \left. \Delta_{i \la j}(t,\xbf_j)
        \right|_{\xbf_j = \xbfhat_{i \la j}(t)},
        \label{eq:QxMapDel}
\eeqa
\end{subequations}
which are the maximum and Hessian of the incoming weak messages.
Since the assumption of the HyGAMP algorithm is that
$\Abf_{ir}$ is small for all weak edges $(i,r)$,
the values of $\xbf_r$ in the maximization
\eqref{eq:DelFnWkij} will be close to $\xbfhat_{i \la r}(t)$.
So, for all weak edges, $(i,r)$,
we can approximate each term $\Delta_{i \la r}(t,\xbf_r)$
in \eqref{eq:HijWk} with the second-order approximation
\beqa
    \lefteqn{ \Delta_{i \la r}(t,\xbf_r) } \nonumber \\
    & \approx &
    \Delta_{i \la r}(t,\xhat_{i\la r}(t))
     - \frac{1}{2} \|\xbf_r -\xbfhat_{i \la r}(t)\|^2_{\Qbf^x_j(t)},
     \label{eq:DelijMaxQuad}
\eeqa
where we have additionally made the approximation $\Qbf^x_{i \la r}(t) \approx
\Qbf^x_r(t)$ for all $i$.
Substituting \eqref{eq:DelijMaxQuad} into \eqref{eq:HijWk},
the maximization \eqref{eq:DelFnWkij} reduces to
\beqa
    \lefteqn{ \Delta^{\rm weak}_{i \ra j}(t,\xbf_j,\zbf_i)
        \approx \mbox{const} } \nonumber\\
        &-& \max_{\substack{\xbf_{\beta(i) \backslash j} \\
        \zbf_i=\Abf_i\xbf} } \left[\frac{1}{2}\sum_{r \in \{\beta(i) \setminus j\}}
        \|\xbf_r - \xbfhat_{i \la r}(t)\|^2_{\Qbf^x_r(t)} \right],
         \label{eq:DelWkQuad}
\eeqa
where the constant term does not depend on $\xbf_j$ or $\zbf_i$.

To proceed, we need to consider two cases separately:
when $j \in \beta(i)$ and when $j \not \in \beta(i)$.
First consider the case when $j \in \beta(i)$.  That is,
$(i,j)$ is a weak edge.  In this case,
a standard least-squares calculation shows that \eqref{eq:DelWkQuad}
reduces to
\beqa
    \lefteqn{ \Delta^{\rm weak}_{i \ra j}(t,\xbf_j,\zbf_i)
        \approx \mbox{const} } \nonumber \\
        &-& \frac{1}{2}
        \|\zbf_i - \Abf_{ij}\xbf_{i \la j}(t) -
        \pbfhat_{i \la j}(t)\|^2_{\Qbf^p_{i \ra j}(t)},
         \label{eq:DelWkQuadij}
\eeqa
where $\pbfhat_{i \ra j}(t)$ and $\Qbf^p_{i \ra j}(t)$ are
given in \eqref{eq:pQpij}.
Also, when $j \in \beta(i)$, the assumption that $\alpha(i)$ and $\beta(i)$
are disjoint implies that $j \not \in \alpha(i)$.  In this case,
$\Delta^{\rm strong}_{i \ra j}(t,\xbf_j, \zbf_i)$ in
\eqref{eq:DelFnStrij} with the objective function \eqref{eq:HijStr}
will not depend on $\xbf_j$, so we can write
\beqa
    \lefteqn{ \Delta^{\rm strong}_{i \ra j}(t,\xbf_j, \zbf_i)
    = \Delta^{\rm strong}_{i}(t,\zbf_i)}\nonumber \\
    &:=& \max_{\xbf }
    \left[ f_i(\xbf_{\alpha(i)}, \zbf_i)
    + \sum_{r \in \alpha(i) }  \Delta_{i \la r}(t,\xbf_r)
    \right], \label{eq:DelStri}
\eeqa
where the maximization is over all $\xbf_r$ for $r \in \alpha(i)$.
Combining \eqref{eq:bpMaxFnPart}, \eqref{eq:DelWkQuadij} and
\eqref{eq:DelStri}, we can write that, for all weak edges $(i,j)$,
\beq \label{eq:DelijG1}
    \Delta_{i \ra j}(t,\xbf_j) \approx G_i(t,\pbfhat_{i \ra j}(t)
        + \Abf_{ij}\xbf_j),
\eeq
where
\beq \label{eq:Gidef}
    G_i(t,\pbfhat_i) := \max_{\xbf_{\alpha(i)},\zbf_i}
    H^z_i(t,\xbf_{\alpha(i)},\zbf_i,
        \pbfhat_i,\Qbf^p_i(t))
\eeq
and $H^z_i(\cdot)$ is defined in \eqref{eq:Hquadi}.

Now define 
$\pbfhat_i(t)$ and $\Qbf^p_i(t)$ as in \eqref{eq:pQpi}.
Using \eqref{eq:pQpiDiff}, neglecting terms of order $O(\|\Abf_{ij}\|^2)$,
and taking the approximation that $\xbfhat_{i\la j}(t) \approx \xbfhat_j(t)$,
\eqref{eq:DelijG1} can be further approximated as
\[  
    \Delta_{i \ra j}(t,\xbf_j) \approx G_i(t,\pbfhat_i(t)
    + \Abf_{ij}(\xbf_j-\xbfhat_j(t))) ,
\]
similar to \eqref{eq:DelijG2}.
Now, similar to \eqref{eq:supdateSp}, let
\begin{subequations} \label{eq:supdate1}
\beqa
    \sbfhat_i(t) &=& \frac{\partial}{\partial \pbfhat}
        G_i(t,\pbfhat_i(t)), \\
    \Qbf^{-s}_i(t) &=& -\frac{\partial^2}{\partial \pbfhat^2}
         G_i(t,\pbfhat_i(t)).
\eeqa
\end{subequations}
Based on the definition of $G_i(\cdot)$ in \eqref{eq:Gidef}
with  $H^z_i(\cdot)$ defined in \eqref{eq:Hquadi}, one can apply
Lemma \ref{lem:GDeriv} to show that \eqref{eq:supdate1} agrees with
\eqref{eq:supdate}.
Using a similar approximation as in the derivation of
the SPA-HyGAMP, one can then obtain the quadratic
approximation in \eqref{eq:DelijQuadWk} for $\Delta_{i \ra j}(t,\xbf_j)$
for all weak edges $(i,j)$.

Next consider the case when $j \not \in \beta(i)$ so that $(i,j)$ is a strong edge.
In this case, $\Delta^{\rm weak}_{i \ra j}(t,\xbf_j,\zbf_i)$ in \eqref{eq:DelWkQuad}
does not depend on $\xbf_j$, so we can write
\beq \label{eq:DelFnWki1}
    \Delta^{\rm weak}_{i \ra j}(t,\xbf_j,\zbf_i)
        \approx  \mbox{const}+ \Delta^{\rm weak}_{i}(t,\zbf_i),
\eeq
where
\beq \label{eq:DelFnWki}
    \Delta^{\rm weak}_{i}(t,\zbf_i)
        := \max_{
        \xbf~:~\zbf_i=\Abf_i\xbf } H^{\rm weak}_{i \ra j}(t,\xbf_{\beta(i)}),
\eeq
with the maximization being over $\xbf$ such that $\zbf_i = \Abf_i\xbf$.
Using a similar least-squares calculation as above,
$\Delta^{\rm weak}_{i}(t,\zbf_i)$ is given by
\beq  \label{eq:DelWkQuadi}
    \Delta^{\rm weak}_{i}(t,\zbf_i)
        := -\frac{1}{2}\|\zbf_i - \pbfhat_{i}(t)\|^2_{\Qbf^p_{i}(t)},
\eeq
and $\pbfhat_i(t)$ and $\Qbf^p_i(t)$ are defined in \eqref{eq:pQpi}.
Combining \eqref{eq:bpMaxFnPart}, \eqref{eq:DelFnWki1}  and
\eqref{eq:DelWkQuadi}, we can write that, for all strong edges $(i,j)$,
\beqa
    \lefteqn{ \Delta_{i \ra j}(t,\xbf_j) \approx \mbox{const} }
    \nonumber \\
    &+& \max_{\zbf_i} \left[ \Delta^{\rm strong}_{i \ra j}(t,\xbf_j,\zbf_i)
    -\frac{1}{2} \|\zbf_i - \pbfhat_i(t) \|^2_{\Qbf^p_i(t)}
    \right] .
    \label{eq:DelFnStrij2}
\eeqa
From \eqref{eq:HijStr} and \eqref{eq:DelFnStrij}, we see that
\eqref{eq:DelFnStrij2} agrees with the factor node update
\eqref{eq:DelijFnOpt} for the strong edges.

We now turn to the variable update steps of the MSA\@.
Since this step is identical to the SPA, one can follow the derivation
in Appendix \ref{sec:SumProdDeriv} to show that
$\Delta_{i \la j}(\tp1,\xbf_j)$ and $\Delta_{i}(\tp1,\xbf_j)$ are given by
\eqref{eq:DelVarijH2} and \eqref{eq:DelVarjH2}, respectively and
$\rbfhat_j(t)$ and $\Qbf^r_j(t)$ are given in \eqref{eq:rupdate}.
Also, the definitions of $\xbfhat_j(t)$ and $\Qbf^x_j(t)$
in \eqref{eq:xhatQxFn} are consistent with
\eqref{eq:xupdate}.

Also, if we let
\[
    \Gamma_j(t,\rbfhat_j) := \argmax_{\xbf_j}
        H_j^x(t,\xbf_j,\rbfhat_{j},\Qbf^r_{j}(t)),
\]
it follows from \eqref{eq:xhatQxFn}, \eqref{eq:DelVarijH2},
and \eqref{eq:DelVarjH2} that
\beqa
    \lefteqn{ \xbfhat_{j}(\tp1)  \approx  \Gamma_j(t,\rbfhat_j(t)) }
        \nonumber \\
    \lefteqn{ \xbfhat_{i \la j}(\tp1) \approx  \Gamma_j(t,\rbfhat_j(t)
        -\Qbf^{r}_{j}(t)\Abf_{i j}^*\sbfhat_i(t)) } \nonumber \\
        &\approx& \xbfhat_j(t)-
    \frac{\partial \Gamma_j(t,\rbfhat_j(t))}{\partial \rbfhat_j}
        \Qbf^{r}_{j}(t)\Abf_{i j}^*\sbfhat_i(t). \label{eq:xhatijGam1}
\eeqa
It can be shown from Lemma \ref{lem:GDeriv} that
\beqan
    \lefteqn{ \frac{\partial \Gamma_j(t,\rbfhat_j(t))}{\partial \rbfhat_j}
    = - \left[ \frac{ \partial^2}{\partial \xbf_j^2}
    H^x_j(t,\xbfhat_j(\tp1),\rbfhat_{j}(t))
    \right]^{-1} \Qbf^{-r}(t) } \nonumber \\
    &\approx& \Qbf^{x}(t)\Qbf^{-r}(t), \hspace{2in}
\eeqan
and hence, from \eqref{eq:xhatijGam1},
\beq \label{eq:xhatijGam2}
    \xbfhat_{i \la j}(\tp1) \approx
    \xbfhat_j(\tp1)-
        \Qbf^x(\tp1)\Abf_{i j}^*\sbfhat_i(t).
\eeq
The proof now follows identically to the derivation of the SPA-HyGAMP\@.

  \section{Derivation of HyGAMP for Group Sparsity} \label{sec:grpSparseUpdate}

This appendix provides a brief explanation of how the
steps in Algorithm~\ref{algo:GAMPGrpSparse} were obtained from SP-HyGAMP in
Algorithm~\ref{algo:GAMP}.  In the description of the SP-HyGAMP algorithm,
we used labels $i$ and $j$ for the factor and variable nodes.
However, the group-sparse estimation problem introduces many other indices.  To avoid
confusion, we adopt the following more explicit (albeit somewhat more cumbersome)
labeling.  The variables nodes will be labeled explicitly by $x_j$ or
$\xi_k$.  For the factor nodes, we use the labels:
\begin{itemize}
\item $a_i$ for the factors $p(y_i|z_i)$;
\item $b_j$ for the factors $P(x_j|\xibf_{\gamma(j)})$; and
\item $c_k$ for the factors $P(\xi_k)$.
\end{itemize}
With this convention, for example, $\Delta_{b_j \la \xi_k}(t,\xi_k)$
represents the message from the variable node $\xi_k$
to the factor node $b_j$ when $j \in G_k$.

Now, in the graphical model in Fig.\ \ref{fig:grpSparse}, the strong edges
are all the edges to the right of the variables $x_j$.  That is, the
strong edges are:
\begin{itemize}
\item between the variables $x_j$ and factors $b_j$ for all $j$;
\item between the variables $\xi_k$ and factors $b_j$ for all $j \in G_k$; and
\item between the variables $\xi_k$ and factors $c_k$ for all $k$.
\end{itemize}
The remaining edges, those between the variables $x_j$ and the factor nodes $a_i$,
are all weak.

With these definitions, we can easily derive the steps in Algorithm~\ref{algo:GAMPGrpSparse} from SP-HyGAMP in Algorithm~\ref{algo:GAMP}.
First, note that all the steps from lines~\ref{line:wkBegin}--\ref{line:gampEnd}
are simply the weak edge updates from Algorithm~\ref{algo:GAMP} specialized to the
case of scalar variables.

To understand the role of the remaining lines,
first consider the message along the strong edge
from the factor node $c_k$ and the variable $\xi_k$.
The factor node $c_k$ corresponds to the prior $P(\xi_k)$ in \eqref{eq:xiProb}.
Since the factor is attached to only one variable node,
the outgoing message in \eqref{eq:DelijFnSp} for this edge reduces to
\beqa \label{eq:delcrx}
    \Delta_{c_k \ra \xi_k}(t,\xi_k) = \log P(\xi_k) = \left\{
        \begin{array}{ll}
        \rho & \mbox{if } \xi_k=1, \\
        1-\rho & \mbox{if } \xi_k=0,
        \end{array} \right.
\eeqa
where the last step follows from \eqref{eq:xiProb}.

Next consider the message along the strong edge
from the variable $\xi_k$ to the factor node
$b_j$ for some $j \in G_k$.
Similar to the case of binary LDPC codes \cite{RichardsonU:09},
since $\xi_k = 0$ or 1, it is convenient to work with log-likelihood ratios
(LLRs).
Given any strong edge between $b_j$ and $\xi_k$, define the LLR,
\beq \label{eq:llrDef}
    \LLR_{j \ra k}(t) := \Delta_{b_j \ra \xi_k}(t,\xi_k=1)-\Delta_{b_j \ra \xi_k}(t,\xi_k=0).
\eeq
The reverse LLR, $\LLR_{j \la k}(t)$ is defined similarly.

Since the variable node $\xi_k$ is not connected
to any weak edges, the variable node output message in \eqref{eq:DelVarStri}
reduces to
\[
    \Delta_{b_j \la \xi_k}(\tp1,\xi_k) = \Delta_{c_k \ra \xi_k}(t,\xi_k)
    + \sum_{i \in \{G_k \setminus j\}} \Delta_{b_i \ra \xi_k}(t,\xi_k).
\]
Therefore the $\LLR$ in \eqref{eq:llrDef} is given by
\beqa
    \lefteqn{ \LLR_{j \la k}(\tp1) = \Delta_{c_k \ra \xi_k}(t,1)-
    \Delta_{c_k \ra \xi_k}(t,0) } \nonumber \\
    & & + \sum_{r \in \{G_k \setminus j\}} \LLR_{r \la k}(t)  \nonumber \\
    &=& \log\left(\frac{\rho}{1-\rho}\right)
    + \sum_{r \in \{G_k \setminus j\}} \LLR_{r \ra k}(t),
\eeqa
where the last step follows from \eqref{eq:delcrx}.

Next consider the message from $b_j$ to $x_j$.  Recall that the factor node $b_j$
corresponds to the distribution $P(x_j|\xibf_{\gamma(j)})$, defined by the variable
$x_j$ in \eqref{eq:xxi}.  Also, this factor node has no weak edges.
Hence, it can be verified that the message
\eqref{eq:DelijFnSp}, propagating from factor node $b_j$ to variable $x_j$, is given by
\beq \label{eq:Delbjrxj}
    \Delta_{b_j \ra x_j}(t,x_j) = \log P_{b_j \ra x_j}(t,x_j),
\eeq
where $P_{b_j \ra x_j}(t,x_j)$ is the probability density 
\beq \label{eq:Pbjrxj}
    P_{b_j \ra x_j}(t,x_j) = \Exp\bigl[ P(x_j|\xibf_{\gamma(j)}) \bigr],
\eeq
and the expectation is over independent variables $\xi_k$ with
\beq \label{eq:xiProbjk}
    P(\xi_k=1) = 1-P(\xi_k=0) = \frac{1}{1+\exp(-\LLR_{j \la k}(t))}.
\eeq
Using the fact that the $P(x_j|\xibf_{\gamma(j)})$ is the conditional distribution
for the variable in \eqref{eq:xxi},
the probability distribution $P_{b_j \ra x_j}(t,x_j)$ in \eqref{eq:Pbjrxj}
can be written
\beq \label{eq:pbjrxjRho}
    P_{b_j \ra x_j}(t,x_j) = P_X(x_j; \rhohat=\rhohat_j(t)),
\eeq
where $P_X(x;\rhohat)$ is the distribution for the variable $X$ in \eqref{eq:xvrho} and
$\rhohat_j(t)$ is the probability
\beqa
    \lefteqn{ \rhohat_j(t) = \Pr\left( \xi_k = 0, \ \forall k \in \gamma(j) \right) } \nonumber \\
    &=& \prod_{k \in \gamma(j)} \frac{1}{1 + \exp(\LLR_{j \la k}(t))}.
    \label{eq:rhojGrp}
\eeqa
Now, the variable node $x_j$ has only one strong edge: that connecting it 
to the factor node $b_j$.  Therefore, the log probability in \eqref{eq:DeljVar} reduces to
\beq \label{eq:DelxjGrp}
    \Delta_{x_j}(\tp1,x_j) = \Delta_{b_j \ra x_j}(t,x_j) -
        \frac{1}{2Q^r_j(t)}|\rhat_j(t)-x_j|^2.
\eeq
Now, as described in equations \eqref{eq:Delbjrxj}
and \eqref{eq:pbjrxjRho},  $\Delta_{b_j \ra x_j}(t,x_j)$ is the log of
the probability distribution for the variable $X$ in \eqref{eq:xvrho} with
$\rhohat(t) = \rhohat_j(t)$.  Hence $\Delta_{x_j}(\tp1,x_j)$ in \eqref{eq:DelxjGrp}
must be the log posterior distribution for the $X$ with the measurement $R=\rhat(t)$
in \eqref{eq:xrGrp}.  Therefore, the expectations and variances in
\eqref{eq:xupdateSp} agree with the expressions in lines \ref{line:xhatGrp} and \ref{line:QxGrp}.

Finally, consider the message from the factor node $b_j$ to a variable node $\xi_k$.
The derivation for this message is similar to the message from $b_j$ to $x_j$.
Specifically, it can be verified that
the factor node message \eqref{eq:DelijFnSp}, applied to the strong edge from
$b_j$ to $\xi_k$, is given by
\beq \label{eq:Delbjrxik}
    \Delta_{b_j \ra \xi_k}(t,\xi_k) = \log P_{b_j \ra \xi_k}(t,\xi_k),
\eeq
where $P_{b_j \ra \xi_k}(t,\xi_k)$ is the probability mass function
\beqa
    \lefteqn{ P_{b_j \ra \xi_k}(t,\xi_k) } \nonumber \\
    &=& \int \exp \Delta_{b_j \la x_j}(\tm1,x_j)
    \Exp\bigCond{ P(x_j|\xibf_{\gamma(j)})}{\xi_k}dx_j, \hspace{0.2in} \label{eq:Pbjrxik}
\eeqa
where the expectation is over independent variables $\xi_k$ with probabilities
in \eqref{eq:xiProbjk}. To evaluate the expectation on the right-hand side of
\eqref{eq:Pbjrxik}, consider the conditional expectation
$\Exp( P(x_j|\xibf_{\gamma(j)})| \xi_k)$.  Since the distribution $P(x_j|\xibf_{\gamma(j)})$
corresponds to the random variable $x_j$ in \eqref{eq:xxi},
\beqa
    \lefteqn{ \Exp\bigCond{ P(x_j|\xibf_{\gamma(j)})}{\xi_k} } \nonumber \\
    &=& \left\{
        \begin{array}{ll}
            P_X(x_j;\rhohat=1) & \mbox{if } \xi_k=1 \\
            P_X(x_j;\rhohat=\rhohat_{j \ra k}(t)) & \mbox{if } \xi_k=0 ,
        \end{array} \right.\label{eq:Pxxi}
\eeqa
where $P_X(x;\rhohat)$ is the probability distribution for the random variable $X$ in
\eqref{eq:xvrho} and
\beqa
    \lefteqn{ \rhohat_{j \ra k}(t) = 1-\Pr\left( \xi_i=0, \ \forall i \in \gamma(j) \setminus k \right) }
        \nonumber \\
    &=& 1 - \prod_{i \in \{\gamma(j) \setminus k\}} \frac{1}{1+\exp\LLR_{i \la k}(t)}.
        \label{eq:rhojkPf}
\eeqa
Also, the edge from variable node $x_j$ to the factor node $b_j$ is the only
strong edge connected to $x_j$.  Therefore, the variable node message \eqref{eq:DelVarStri}
applied to that edge reduces to
\beq \label{eq:Delbjlxj}
    \Delta_{b_j \la x_j}(\tm1,x_j) = -\frac{1}{2Q^r_j(\tm1)}|x_j-\rhat_j(\tm1)|^2.
\eeq
Substituting \eqref{eq:Pxxi} and \eqref{eq:Delbjlxj} into \eqref{eq:Pbjrxik}
we obtain that
\beqa
    \lefteqn{ P_{b_j \ra \xi_k}(t,\xi_k) } \nonumber \\
    &\propto& \left\{
        \begin{array}{ll}
            p_R(\rhat_j(\tm1); Q^r_j(\tm1), 1)  & \mbox{if } \xi_k=1 \\
            p_R(\rhat_j(\tm1); Q^r_j(\tm1), \rhohat_{j \ra k}(t) & \mbox{if } \xi_k=0,
        \end{array} \right.    \label{eq:PbjrxikRho}
\eeqa
where $p_R(r;Q^r,\rhohat)$ is the probability distribution of the scalar random variable
$R$ in \eqref{eq:xrGrp} with $X$ being distributed in \eqref{eq:xvrho}.
The LLR corresponding to \eqref{eq:PbjrxikRho} is thus given by
\beqan
    \lefteqn{ \LLR_{j \ra k}(t) = \log P_{b_j \ra \xi_k}(t,\xi_k=1) } \nonumber \\
    & & - P_{b_j \ra \xi_k}(t,\xi_k=0) \nonumber \\
        &=& \log p_R(\rhat_j(\tm1); Q^r_j(\tm1), \rhohat=1) - \nonumber \\
        & & - \log p_R(\rhat_j(\tm1); Q^r_j(\tm1), \rhohat=\rhohat_{j \ra k}(t)),
\eeqan
which agrees with \eqref{eq:llrjrk}.

\else
  
  \blue
  
  \black
\fi

\bibliographystyle{IEEEtran}
\bibliography{../bibl}

\end{document}